\documentclass[twocolumn,final,numbook,nospthms]{svjour3}

\usepackage{amsmath}
\usepackage{hyperref}
\usepackage{cleveref}
\usepackage{mathrsfs}
\usepackage{graphicx}
\usepackage{epsfig,bm,epsf,float}
\usepackage{hhline}
\usepackage{txfonts,makecell}
\usepackage[numbers]{natbib}
\usepackage{xcolor}

\crefname{equation}{eq.}{eqs.}

\newcommand{\diff}{\mathrm{d}}

\def\be{\begin{equation}}
\def\ee{\end{equation}}
\def\kmsMpc{\ensuremath\,\rm{km}\,\rm{s}^{-1}/\rm{Mpc}}

\newcounter{mysfig}
\renewcommand\themysfig{(\alph{mysfig})}
\makeatletter
\newcommand\Scaption[1]{%
\refstepcounter{mysfig}%
\vskip.5\abovecaptionskip
  \sbox\@tempboxa{\small\themysfig~#1}%
  \ifdim \wd\@tempboxa >\hsize
    \small\themysfig~#1\par
  \else
    \global \@minipagefalse
    \hb@xt@\hsize{\hfil\box\@tempboxa\hfil}%
  \fi
  \vskip\belowcaptionskip}
\makeatother

\begin{document} 

  \title{Cosmological constraints from low-redshift data}
  
  \author{
    Vladimir V. Lukovi\'{c}
    \and Balakrishna S. Haridasu
    \and Nicola Vittorio
  }
  
  \institute{
    V. V. Lukovi\'{c}\and N. Vittorio\at Dipartimento di Fisica, Universit\`{a} di Roma "Tor Vergata", Via della Ricerca Scientifica 1, I-00133, Roma, Italy\\Sezione INFN, Universit\`{a} di Roma "Tor Vergata", Via della Ricerca Scientifica 1, I-00133, Roma, Italy\\\email{vladimir.lukovic@roma2.infn.it,\,nicola.vittorio@roma2.infn.it}\and
    B. S. Haridasu \at Gran Sasso Science Institute, Viale Francesco Crispi 7, I-67100 L'Aquila, Italy\\\email{sandeep.haridasu@gssi.it}
  }
  
%  \titlerunning{text}
  
  \authorrunning{V. V. Lukovi\'{c} et al.}

  \maketitle

  \tableofcontents{}

  
  \begin{abstract}
    In this paper we summarise the constraints that low-redshift data --such as supernovae Ia (SN Ia), baryon acoustic oscillations (BAO) and cosmic chronometers (CC)-- are able to set on the concordance model and its extensions, as well as on inhomogeneous but isotropic models.
    We provide a broad overlook into these cosmological scenarios and several aspects of data analysis.
    In particular, we review a number of systematic issues of SN Ia analysis that include magnitude correction techniques, selection bias and their influence on the inferred cosmological constraints.
    Furthermore, we examine the isotropic and anisotropic components of the BAO data and their individual relevance for cosmological model-fitting.
    We extend the discussion presented in earlier works regarding the inferred dynamics of cosmic expansion and its present rate from the low-redshift data.
    Specifically, we discuss the cosmological constraints on the accelerated expansion and related model-selections.
    In addition, we extensively talk about the Hubble constant problem, then focus on the low-redshift data constraint on $H_0$ that is based on CC.
    Finally, we present the way in which this result compares the two of high-redshift $H_0$ estimate and local (redshift zero) measurements that are in tension.
  \end{abstract}

  \section{Introduction}
    \label{sec:int}
    The current cosmological paradigm has been well developed over decades of severe scrutiny with general relativity (GR) as the underlying framework.
    A century after its first proposal \cite{Einstein15}, GR still reigns as the most complete description of gravity, also owing to the recent observations of gravitational waves \citep{Abbott17a}.
    In this era of precision cosmology, the observations of ``high-redshift'' cosmic microwave background radiation (CMB) \citep{WMAP9,Planck15} and several ``low-redshift'' observables provide us with very high constraining power on the cosmological parameters.
    While the technical advancements in the observational cosmology help us to acquire a wealth of data, we are yet unable to formulate a theoretical model with well understood components, i.e., content of the universe.
    The dark sector of the Universe comprises of non-luminous dark matter (DM) component and a mysterious dark energy (DE) component.
    The latter is pivotal to explain the late-time acceleration phase of the cosmic expansion, which was for the first time directly confirmed with the observations of type Ia supernovae by \citet{Riess98} and \citet{Perlmutter99}, and further supported with more recent SN Ia data \citep{Suzuki12,Betoule14}.
    Other low redshift data, including cosmic chronometers (CC) \citep{Jimenez02} and the baryon acoustic oscillations (BAO) \citep{Eisenstein05} also support an accelerating universe.
    As an independent observation, the CMB data from \citet{Planck15} (hereafter P15) are in excellent agreement with these results and they provided the most stringent constraints on the cosmological models.
    Interestingly, Planck mission provided the temperature anisotropy map at the precision of cosmic-variance.
    In fact, the current and future CMB experiments are focused on polarisation maps and secondary anisotropies (see e.g. \citep{Core17,ACT13}).
    
    The low-redshift data has undergone several important improvements in the last decade.
    As the SN Ia compilations  gather more and more SN, it demands for a more robust statistical method to compile a homogeneous dataset that can be implemented to test cosmological models.
    The most used compilation has been provided as the joint light-curve analysis (JLA) dataset in \citet{Betoule14}, with 740 selected SN Ia up to redshift $z \lesssim 1.4$.
    On the other hand, the BAO data has undergone its fair share of advancement with the latest Sloan digital sky survey (SDSS) data release in \citet{Alam17}.
    The observations of BAO feature in the cross-correlated Lyman-$\alpha$ spectrum at $z\sim2.4$ \citep{Delubac15,Font-Ribera13} have complemented the galaxy-clustering data at lower redshifts to provide more stringent constraints and the ability to distinguish the dynamics of different cosmological models \citep{Haridasu17a}.
    Complementing these two stringent datasets and CC, which provide measurements of the expansion rate at different redshifts, is a powerful tool for estimating the Hubble constant value- $H_0$ \citep{Lukovic16}.

    As a matter of fact, the Hubble constant is an important cosmological parameter that defines the extragalactic distance scale, through the Hubble radius $d_H=c/H_0$.
    It can be constrained both from CMB measurements and low-redshift data.
    The estimate of $H_0$ from high-redshift data is strongly correlated to the cosmic expansion history, and hence, the presumed theoretical model used in data analysis.
    Similarly, the Hubble constant value obtained from the low-redshift data is specifically correlated to the late-time dynamics of the cosmic expansion.
    Therefore, there have been many attempts to provide an independent local measurement of the Hubble constant that does not rely on the theoretical model of the cosmic expansion (for a review see \citet{Freedman10,Freedman17}).
    Unfortunately, the most stringent CMB-based estimate by \citet{Planck16}, $H_0=66.93\pm0.62\kmsMpc$ is in tension with the most precise direct measurement by \citet{Riess18}, $H_0= 73.45 \pm 1.66\kmsMpc$.

    Besides a number of open problems, the standard model of cosmology is very successful in explaining all observations.
    Nevertheless, the increasing amount of cosmological data begs for more robust statistical methods to test the same and other theoretical possibilities.
    Among many others, scalar field DE models and modified gravity scenarios have been the most sought out alternatives to the standard model, to attribute a physical nature to the late-time acceleration phase.
    In this respect we have studied several extensions of the standard model using the low-redshift data, which are crucial to describe the late-time dynamics of the universe and provide strong evidence for the accelerated expansion \citep{Haridasu17}.
    In addition, we also consider a different class of models, based on the Lema\^{i}tre-Tolman-Bondi (LTB) metric, which describes an isotropic but inhomogeneous Universe \citep{Krasinski97}, to stress the dependence of the Hubble constant estimates on the assumed theoretical model.
    
    In this article we briefly summarise our work using the low-redshift datasets to test cosmological models, presented in \citet{Lukovic16,Haridasu17,Haridasu17a}.
    We have progressively taken into account the contemporary discussions regarding the SN Ia analysis and the ever most recent BAO data.
    We begin by providing a broad overlook into the cosmological scenarios in \Cref{sec:th}, then we review the SN Ia analysis in \Cref{sec:SN}, followed by a brief introduction to the BAO dataset in \Cref{sec:BAO}.
    In \Cref{sec:CC} we address the direct measurements and the model-dependent methods of estimating the present cosmic expansion rate.
    The resulting constraints on the cosmological models and on the present expansion rate are discussed in \Cref{sec:res} and \Cref{sec:H0}, respectively.
    Finally, we summarise our findings in \Cref{sec:con}.
    Throughout the article, the values of cosmic expansion speed, e.g. $H_0$, are always quoted in assumed units of $\kmsMpc$.

  \section{Theoretical scenarios}
    \label{sec:th}
    The story of the cosmological constant $\Lambda$ starts 100 years ago, following the first attempts to construct a relativistic model of the Universe.
    The original idea for $\Lambda$ was very different from the way in which it is used today.
    It was personal conviction of \citet{Einstein17} that the Universe should be static and eternal, which gave rise to the introduction of an additional term in his field equations (EE) of general relativity:
    \be
      {G^\mu}_\nu-\Lambda{g^\mu}_\nu=\kappa {T^\mu}_\nu
      \label{eq:EE}
    \ee
    The cosmological constant multiplies the metric tensor ${g^\mu}_\nu$ and, therefore, this term satisfies the covariant zero-divergence, with the energy-momentum tensor (${T^\mu}_\nu$) and the Einstein tensor (${G^\mu}_\nu$).
    Einstein arrived to \Cref{eq:EE} on the basis of symmetry, trying to relate the geometry of the space-time from one side to the matter-energy content on the other side, via only a proportionality constant $\kappa$. 
    Although the additional term with cosmological constant is permitted as it is following the symmetry of energy-momentum tensor and EE \citep{McCrea71,Lovelock71,Lovelock72}, the physicists have been ever since puzzled with the underlaying nature of this term (for a historical review see \citet{Raifeartaigh17}).
    At the beginning, Einstein assumed a specific negative value for $\Lambda$ that counterparts the effect of the mass-energy content, in order to construct a static metric and the model of the eternal Universe.
    As it often happens in the development of science, the observations of \citet{Hubble29} provided evidence for the cosmic expansion, contrary to Einstein's expectation (see also \citet{Lemaitre27}). 
    This lead to reduced interest in the model of the Universe with cosmological constant, until only three decades ago \citep{Bardeen87,Efstathiou90,Bahcall92}.
    %Among the glorious cosmological observations from the end of last century were also that of high-redshift supernovae of type Ia (SN Ia) which revealed the present accelerating phase of the cosmic expansion \citep{Riess98,Perlmutter99}.
    Not long after, the observations of high-redshift supernovae of type Ia (SN Ia) revealed the present accelerating phase of the cosmic expansion \citep{Riess98,Perlmutter99}.
    While the existence of dark matter component is well accepted phenomena (see e.g. \citet{Lukovic14}), the accelerated expansion is not expected in a universe consisting of baryons and cold dark matter alone.
    This strongly supported the need for an extra component, today referred to as \textit{dark energy}.
    
    Adding a cosmological constant term to the left hand side of \Cref{eq:EE}, as Einstein did, has the same effect as adding a covariant energy density with negative pressure to the right hand side. 
    This early interpretation of cosmological constant was given in 1918 by \citet{Schroedinger18}, right after Einstein's (see also \citet{Harvey12}).
    Much later the mysterious energy density was suggested to be the vacuum energy \citep{Zeldovich67,Sarkar07}, but that was only a beginning.
    The search for the physical interpretation of dark energy, which still evades our understanding, has led to a cascade of theoretical scenarios.
    Field theory provides an ideal framework for describing DE as a dynamical fluid.
    Scalar fields with slowly varying potentials \citep{Ratra87,Peebles87,Wetterich88,Tsujikawa13}, or models in which the acceleration is driven by the kinetic energy of scalar fields \citep{Chiba00} belong to this class.
    Alternatively, one may consider that gravity itself is weakened on large scales, i.e. that there is a low-energy modification to general relativity (GR) which would effectively produce the late-time acceleration \citep{Clifton12}.
    Although possible, such GR discrepancies are difficult to test experimentally.
    Degeneracies in different theoretical scenarios led to the difficulty of choice, encouraging the development of methods for model selection \citep{Mukherjee06}.
    Among all, the simplest homogeneous and isotropic model with positive cosmological constant and pressureless matter content ($\Lambda$CDM) stands out, as it remarkably fits in concert with both CMB and low-redshift observations.
    Commonly known as \textit{standard model in cosmology}, $\Lambda$CDM is also very well supported by the predictions of baryon matter content from primordial nucleosynthesis.
    The extensive matter distribution surveys available today are fundamental to establish the late time evolution of the universe, necessary to distinguish theoretical models based on their dynamics.
    This challenging task is the most sought out problem in cosmology today.

    A study of properties of the Universe through the use of astrophysical data available today, first of all, requires the theoretical expectations for relevant astrophysical observables.
    The following subsections present a number of different cosmological models with viable explanations for late-time evolution and provide analytical expressions for the cosmic expansion rate and distance measures.
    In our work, we adopt the approach of parametrising relevant physical quantities using standard methods for model-fitting and data analysis.

    \subsection{Concordance model of cosmology and its extensions}
    \label{ssec:FLRW}
    The first practical solution for cosmological metric that satisfies the EE was found by \citet{Friedmann22,Friedmann24} and subsequently further developed \citep{Lemaitre33,Robertson35,Walker37}. The FLRW metric describes a homogeneous and isotropic cosmic fluid:
    \be
      \diff s^2=c^2 \diff t^2-R^2(t)\left(\frac{\diff r^2}{1-k r^2}+r^2\diff\Omega^2\right)
      \label{eq:1}
    .\ee
    Here $R(t)$ is a scale factor in units of length, while the reduced scale factor can also be introduced as $a(t)=R(t)/R_0$, relative to the value at the present age of the Universe, $R_0=R(t_0)$. 
    The curvature parameter $k=-1,\ 0,\ +1$ characterises the open, flat, and closed geometry, respectively. 
    Assumptions of homogeneity and isotropy together embody the \textit{cosmological principle} and represent the building blocks of the standard model of cosmology.
    In this framework, the cosmic expansion rate, or Hubble parameter, is defined as
    \begin{equation}
      H(z) \equiv {1\over R}{\diff R\over \diff t} = -{1\over1+z}{\diff z\over \diff t}
      \label{eq:defH}
    ,\end{equation}
    which is a function of redshift $z$.
    
    The dynamical properties of a cosmic fluid depend on its contents. 
    For the study of late-time cosmic evolution, relevant constituents of the cosmic fluid are baryons, dark matter and dark energy.
    Each component is characterised with pressure $p$, density $\rho$ and equation of state (EoS) parameter $w\equiv p/\rho$.
    While baryons and dark matter are both considered to be cold and pressureless ($w\equiv0$) during this epoch, the dynamical features of dark energy fluid have been a matter of discussion \citep{Copeland05,Bahamonde17}.
    A simple and natural framework for the study of late-time dynamical expansion and cosmological constant problem is given by scalar fields.
    The first consideration of scalar fields in cosmology was proposed by \citet{Guth81} to explain the inflationary paradigm.
    Later they were extensively explored for various purposes, among which is the nature of dark energy.
    The DE field models are generally assumed with a background FLRW metric.
    There exists a wealth of literature in this context (for review see e.g. \citet{Copeland06,Li11}).
    As already mentioned, the cosmological constant can be interpreted as DE fluid with negative pressure, i.e. $w=-1$.
    The key change in dynamical dark energy models is that the pressure and energy density have freedom to evolve in time, and hence, have a varying EoS parameter $w_{\phi}(z)$, defined by the nature of the scalar field.
    Quintessence field models are the simplest among them.
    Inspired from the inflationary models, they are characterised by a canonical scalar field that is minimally coupled to gravity \citep{Caldwell98,Zlatev99,Tsujikawa13}.
    However, many alternative classes of non-canonical fields have been proposed: $k$-essence models, whose kinetic term gives rise to late time acceleration \citep{Chiba00,Armendariz01,Armendariz00}; tachyon fields \citep{Bagla03,Padmanabhan02}; interacting DE \citep{Cohen99,Amendola00,Cai05}; early DE models \citep{Steinhardt99,Doran01}, which try to alleviate some conceptual problems of cosmological constant; Chaplygin gas models \citep{Kamenshchik01,Pun08,Deng11,Pourhassan14}; phantom DE field, which have $w<-1$\citep{Caldwell02,Caldwell03,Carroll03} et cetera.
    
    The content of the Universe is related to its expansion rate through the Friedmann equation:
    \begin{equation}
      H(z) = H_0\sqrt{\Omega_m(1+z)^3  + \Omega_k(1+z)^2 +\Omega_{DE}f(z)}
      \label{eq:Feq1}
    ,\end{equation}
    where $H_0\equiv H(z=0)$, while $\Omega_m$ and $\Omega_{DE}$ represent the density of the total matter and dark energy components in units of the critical density $\rho_c=3H_0^2/(8\pi G)$. 
    The remaining term $\Omega_k\equiv-k c^2/\left(H_0 R_0\right)^2$ satisfies the closure equation $\Omega_k = 1-\Omega_m-\Omega_{DE}$. 
    The contribution of dark energy term is defined with the function
    \be
      f(z) = \exp\left( 3 \int^{z}_0 \frac{1+w(\xi)}{1+\xi} \diff \xi\right)
    .\ee
    The case of constant EoS parameter $w\equiv-1$ and $\Omega_{DE}=\Lambda c^2/(3H_0^2)$ represents the flat $\Lambda$CDM model of the Universe.
    Likewise, the flat $w$CDM model considers the dark energy fluid with free but constant EoS parameter $w\neq-1$, hence different from the concordance model.
    Indeed, all $\Lambda$CDM extensions have more model parameters characterising them.
    For this reason, models that study the dynamics of dark energy usually keep the assumption of flat geometry $\Omega_k=0$, effectively reducing the number of free parameters.
    Such an assumption is also justified by the curvature constraints based on observations of CMB temperature anisotropies \citep{WMAP9,Planck15}.
    Nevertheless, curvature contribution can also be studied by considering the free curvature parameter $\Omega_k\neq0$.
    We note these models as, e.g., $k\Lambda$CDM and $kw$CDM.
    
    While each form of the DE Lagrangian and specifically of the field's potential produce different dynamical evolutions, the constraining power of the current data is limited and, hence, it is preferable to control the number of the degrees of freedom in data analysis.
    To this end, cosmologists consider analytic parameterisations for $w(z)$ function with only a few free parameters \citep{Linder05,Jassal05,Feng12,Yang17}.
    Among these, perhaps the simplest and the most used one is a first order Taylor expansion of the EoS parameter with respect to (w.r.t.) the scale factor today, proposed by \citet{Chevallier01,Linder03a}.
    Also known as the CPL parametrisation, it reads:
    \be
      w(z)=w_0+w_a (1-a)=w_0+w_a {z\over1+z}
      \label{eq:CPL}
    \ee
    Flat model with this dynamic EoS parameter we note as $w_0w_a$CDM.
    
    An alternative explanation for late-time accelerated expansion is that the underlaying laws of gravity are governed not by GR, but by a different theory.
    Plenty of distinctive, considerable scenarios have been developed for exploring this possibility, broadly called \textit{modified gravity} models \citep{Sotiriou10,Joyce15,Clifton12}.
    Since parameterised DE models are simply examining the deviation from the standard $\Lambda$CDM model, this approach can be effectively used for studying the dynamics of cosmic expansion in the most of modified gravity models as well \citep{Linder03b,Alam03}.
    However, the physical interpretation of the cosmological parameters would not be the same as in DE models.
    Perhaps the only way to phenomenologically disentangle pure DE from the modified gravity models is at medium cosmic scales, e.g. by observing the formation and evolution of the large-scale structures \citep{Joyce16}.
    
    The wealth of astrophysical data in modern age is enabling us to constrain different cosmological distance measures.
    For instance, observations of SN Ia provide us with the measurement of luminosity distance $d_L$.
    Contrastingly, the observations of baryon acoustic peak in matter power spectrum constrain the transverse (angular) comoving distance, $d_M$, together with the Hubble parameter, $H$.
    However, BAO data are also frequently reported in terms of the comoving volume averaged angular diameter distance, $d_V$, and the \citet{Alcock79} parameter, $AP$.
    In models based on FLRW metric, these distance measures are calculated according to the equations:
    \begin{align}
      d_M(z) &= \frac{c}{\sqrt{\Omega_k}}\sinh\left(\sqrt{\Omega_k}\int_0^z\frac{d\xi}{H(\xi)}\right)
      \label{eq:dM}\\
      d_L(z) &= (1+z)d_M(z)
      \label{eq:dL}\\
      d_{V}(z) &= \left[d_{M}^{2}(z)\frac{c z}{H(z)}\right]^{1/3}
      \label{eq:dV}\\
      AP(z) &=  d_{M}(z) \frac{H(z)}{c}
    \end{align}
    Often we use the dimensionless variables as: 
    \begin{align}
      D_M(z) &= d_M(z) / r_d\\
      D_V(z) &= d_V(z) / r_d\\
      \intertext{in BAO analysis, and }
      D_L(z) &= d_L(z) / d_H
    \end{align}
    in SN Ia analysis.
    Here $r_d$ (usually $\sim$150\,Mpc) is the sound horizon at the drag epoch and $d_H = c / H_0 \simeq 3000 h^{-1}$Mpc is the Hubble radius.
    The presented formulae are used for fitting low-redshift data to cosmological models based on FLRW metric.

    \subsection{Power-law cosmologies}
    A power-law cosmological model has been a well sought out alternative framework to the standard Friedmann cosmology owing to its simplistic modelling and ease to test against the data.
    In a flat, power-law cosmological model the scale factor evolves in time as $a(t) \propto t^{n}$ (t being the proper time), with the Hubble equation $H(z) = H_{0}(1+z)^{1/n}$.
    It has been shown in \citet{Dolgov97} that classical fields non-minimally coupling to spacetime curvature can give rise to a back-reaction from singularities, which can change the nature of expansion from exponential to power-law.
    %This model has also been discussed as an alternative to $\Lambda$CDM model in order to solve the horizon problems in \citet{Sethi05}, especially for $n < 1$.
    %Naturally, as these models do not constrain the matter content of the universe, the flatness problem is overcome.
    In the context of acceleration, power-law cosmologies with $n \geq 1$ have been explored against data in several works such as \citep{Gehlaut03, Dev08, Zhu08, Dolgov14, Rani15}, finding $n \sim 1.5$ consistently.
    
    Another class of models that have been of interest are the linear coasting models \citep{John00, Dev02}.
    Such a model can be obtained as a specific case of the power-law cosmology with $n=1$.
    In the more recent literature, these models have also been studied under the alias as the $R_h = ct $ models \citep{Melia12}.
    A number of works \citep{Melia14,Melia16} argue that $R_h = ct $ model is statistically preferred over the standard $\Lambda$CDM model.
    On the contrary, several other works \citep{Bilicki12,Shafer15,Tutusaus16,Haridasu17,Lonappan17}, using more recent data and in particular joint-analysis of several low- and high-redshift data, have shown that the linear and power-law models are highly disfavoured, when compared with the standard model.
    Furthermore, \citet{Lewis16} claim that a model like $R_h=ct$ gives rise to an extended period of primordial nucleosynthesis which in turn results in lower abundances of light elements.

    The linear coasting models are often misinterpreted (c.f. \citep{Mitra14,Bilicki12}) as the empty Milne model \citep{Milne35} with $\Omega_m = \Omega_{\Lambda} = 0$, as the functional form of the Hubble expansion rate is exactly the same for the two.
    However, the Milne model describes an empty universe with the curvature parameter $\Omega_k =1$, satisfying the closure equation.
    Whereas, the $R_h = ct$ model is generally considered as a flat model with $\Omega_k = 0$, but could also have arbitrary curvature.
    Hence, the geometrical distance measures in these two models are clearly not the same, and the two models perform differently when compared against the data \citep{Haridasu17}.
    Although coinciding with the power-law model for $n=1$, the flat $R_{h}=ct$ model is often physically interpreted as the FLRW model with a constant total EoS parameter $w_{tot}=-1/3$ \citep{Melia12}.
    It is worthwhile noting that, more generally, the standard Friedmann \Cref{eq:Feq1} reduces to the functional form of any power-law model for selected parameter values: $\Omega_{m}=0$, $\Omega_{DE}=1$ and
    \begin{equation}
      \label{eqn:wn}
      w = \frac{2-3n}{3n}.
    \end{equation}
    The power-law and $R_h = ct$ models are unable to provide any further understanding for the content of the universe, which also makes it harder to study the structure formation and physics of early universe.
    However, the presented functional equivalences can be conveniently used to compare and discuss the data constraints inside of the $w$CDM model itself.

    \subsection{Challenging the cosmological principle: LTB models}
    Soon after the Friedmann solution of the EE, the cosmological metric was extended to a more general solution describing the isotropic, but inhomogeneous fluid. 
    It was found by \citet{Lemaitre33}, and later developed further by \citet{Tolman34,Bondi47}. 
    In the case of a pressureless cosmic fluid, this solution, known as LTB metric (see also \citet{Krasinski97}), can be written as
    \be
      \diff s^2=c^2 \diff t^2-\frac{{R'}^2(r,t)}{1-k(r)}\diff r^2-R^2(r,t)\diff\Omega^2,
      \label{eq:9}
    \ee
    Here $k(r)$ determines the spatial curvature of 3D space, while the derivatives w.r.t. comoving radial coordinate $r$ and w.r.t. time $t$ are denoted using prime and dot symbols, respectively. 
    As in the case of FLRW metric, $R(r,t)$ is the scale factor in units of length, and we can introduce the reduced scale factor as well: $a(r,t)\equiv R(r,t)/R_0(r)$, with $R_0(r)\equiv R(r,t_0)$.
    
    As mentioned above, the reason for introducing dark energy into the cosmological models was to explain the present accelerated state of the Hubble flow confirmed firstly by the SN Ia observations.
    However, in inhomogeneous cosmological model, the functional form of the cosmic expansion w.r.t. redshift ($H(z)$) can have the same feature of (apparent) acceleration at low redshifts not due to the presence of a component with negative pressure, but due to a gradient in the matter density profile.
    Although such a cosmological model does not need dark energy to explain the apparent acceleration, it would mean that we are positioned at a special place in the Universe - in a giant underdense region, also called the \textit{void}.
    The bare existence of a giant void is going against the cosmological principle.
    In principle, the cosmic isotropy is well confirmed from CMB observations, while the tests of homogeneity are not as easy to perform, also because of difficulty to differ temporal and spatial evolution on the past light cone, e.g. \citep{Maartens11,Clarkson12,Keenan13,Whitbourn14,Bohringer15}).
    Following the findings from high-redshift SN Ia, the void models gained popularity as the alternative explanation for cosmic acceleration \citep{Celerier00,February10,Nadathur11,Sundell15}.
    
    Owing to inhomogeneity of the cosmic fluid, all its parameters such as pressure, density and EoS parameter $w$ may depend not only on time, but also on the radial coordinate, introducing a more complex picture.
    In addition, the LTB model is characterised with two expansion rates, the radial and the transverse one:
    \be
      H_\parallel(r,t)\equiv\dfrac{\dot{R}'(r,t)}{R'(r,t)}\,\,,\hspace{1.5em}H_\perp(r,t)\equiv\dfrac{\dot{R}(r,t)}{R(r,t)}=\dfrac{\dot{a}(r,t)}{a(r,t)}
      \label{eq:Hpp}
    \ee
    The local expansion rate at the present time is given by $ H_0\equiv H_\parallel(r=0,t_0)=H_\perp(r=0,t_0)$.
    LTB models often consider a cosmic fluid composed of a pressureless inhomogeneous cold matter (with $w=0$) and, possibly, a dark energy fluid (with EoS parameter e.g. $w=-1$). 
    Such a DE fluid component is equivalent to having an inhomogeneous $\Lambda$ term that is constant in time, but with a different value for each sphere $\Lambda(r)$.
    Recalling the Jebsen-Birkhoff theorem \citep{Deser05}, we expect that each shell evolves independently of the others, as a FLRW model with the same values of the fluid parameters.
    Hence, solving EE leads to the analogue of Friedmann equation for the expansion rate of a shell (for detailed derivation see \citet{Enqvist07,Nadathur11,Lukovic16})
    \begin{equation}
      {H^2_\perp(r,t)}=H_0^2(r)\left[\dfrac{\Omega_m(r)}{a(r,t)^3}+\dfrac{\Omega_k(r)}{a(r,t)^2}+\Omega_\Lambda(r)\right]
      \label{eq:LTBF}
    \end{equation}
    As in the FLRW model, $\Omega_m(r)\equiv8\pi G\rho_m(r)\left/3H_0^2(r)\right.$, $\Omega_\Lambda(r)\equiv\Lambda(r)c^2\left/3H_0^2(r)\right.$ and $\Omega_k(r)\equiv-k(r) \left/H_0^2(r) R_0^2(r)\right.$ are rescaled densities in units of the critical density. 
    For each shell we have $\Omega_m(r) + \Omega_k(r) + \Omega_\Lambda(r)\equiv 1$. In the special cases of $\Omega_\Lambda(r)\equiv0$ or $\Omega_k(r)\equiv 0$, the \Cref{eq:LTBF} can be integrated analytically, while in general has to be solved numerically.
    
    LTB models can have various matter density profiles, but probably the simplest and the most studied one is the Gaussian profile:
    \be
      \Omega_m(r)=\Omega_{\rm out}-(\Omega_{\rm out}-\Omega_{\rm in})e^{-r^2/2\rho^2}
      \label{eq:Gaussian}
    \ee
    Here $\Omega_{\rm in}\leq\Omega_{\rm out}$ are the density parameters at the centre and outside of this underdense region, while the parameter $\rho$ defines its size.
    The value of $\Omega_{\rm out}$ is fixed to the asymptotic background homogeneous model in order to be consistent with the inflationary paradigm.
    One can see that $r\to\infty$ recovers FLRW model with $\Omega_m=\Omega_{\rm out}$.
    
    Only the observer located at the very centre of the void will enjoy the isotropic view of the Universe, while the off-centre position inside the void will introduce a level of anisotropy. The angular diameter distance for the on-centre observer \citep{Ellis07} is isotropic and equal to
    \begin{equation}
      d_A(z)=R(r(z),t(z))\ .
      \label{eq:distanceon-centre}
    \end{equation}
    Since the Etherington duality relation $d_L = (1 + z)^2 d_A$ remains valid in inhomogeneous models \citep{Kristian66}, we have
    \begin{equation}
      d_L(z)=(1+z)^2 R\left[r(z),t(z)\right]\ .
      \label{eq:distanceon-centre}
    \end{equation}
    This is obviously in a privileged position and against the Copernican principle. 
    The off-centre location of the observer gives rise to some complication of the mathematical formalism (see \citet{Alnes06b}), but all the relevant observables can be derived.

  \section{Type Ia supernovae}
    \label{sec:SN}
    Since the discovery of stellar parallax \citep{Bessel44} and period-luminosity relation of Cepheids \citep{Leavitt08}, measuring distances to astrophysical objects on the celestial sphere has remained a very difficult task.
    The most reliant objects for measuring cosmological distances up to date are supernovae Ia.
    Specifically, the observations of high-redshift SN Ia most prominently influenced the development of concordance model and especially the cosmological constant dilemma \citep{Filippenko05}.
    Although among cosmologists they have risen to popularity by being the best known high-redshift standard candles, today we know that SN Ia are not \textit{perfect} standard candles.
    Luckily, the variation of their luminosities can be further reduced with standardisation techniques.
    The methods for standardising SN Ia are based on observed phenomenology and correlations between their intrinsic parameters.
    However, the theoretical aspects are not completely understood \citep{Nobili08,Bronder08,Holwerda15}.
    This led to various discussions about the statistical analyses of SN Ia and the quality of cosmological inferences derived from SN Ia datasets.
    
    In this section we chronologically revise the improvements in both the data available and the standardisation methods in use, as well as their effects on the cosmological constraints.

    \subsection{From dozens to a thousand SN Ia}
    The first big projects dedicated to the discovery and systematic observation of SN Ia at high redshifts started during the last decade of the XX century.
    Two teams - one led initially by Brian P. Schmidt and then by Adam Riess, and the other one led by Saul Perlmutter - provided large enough samples of high-redshift SN Ia for the first significant evidence of an accelerated cosmic expansion.
    \citet{Riess98} used 16 high-redshift and 34 nearby SN Ia to place constraints on the cosmological parameters: the data favour a current accelerated state of the Universe at more than $3\sigma$ confidence level, implying a positive value for the cosmological constant.
    Combining the original observations of 42 high-redshift SN Ia with the Cal\'{a}n/Tololo Supernova Survey at redshifts below 0.1, \citet{Perlmutter99} found $\Omega_\Lambda=0.7\pm0.1$ for flat $\Lambda$CDM model.
    Since then, the Supernova Cosmology Project, co-founded by Perlmutter, contributed to the largest SN datasets, such as Constitution (397 SN from \citet{Hicken09}), Union2.1 (580 SN drawn from 19 datasets, see \citet{Suzuki12}), and finally the Joint Light-curve Analysis (JLA) dataset that unifies measurements of 4 major subsamples in a joint statistical analysis \citep{Betoule14} (hereafter B14).
    Consisting of 740 selected SN Ia, JLA dataset was carefully implemented to overcome the most important limitations identified in earlier analyses, and were ever since used for testing cosmological models and studying the dark energy paradigm.
    Nevertheless, the number of observed SN Ia still rises as the most recent \textit{Pantheon Sample} has 1049 selected SN Ia ranging from $0.01 < z < 2.3$ \citep{Scolnic17,Jones17}.
    
    \subsection{Standardisation techniques}
    The supernova explosion happens extremely fast on the astronomical scales, but it is not an immediate process. 
    Within a period of weeks the SN Ia luminosity changes by several magnitudes, reaching the peak of the lightcurve, decreasing and then fade more steadily.
    The accepted model of SN Ia is that they arise from a thermonuclear explosion of a white dwarf star, triggered by reaching the maximum mass of the degenerate core - the Chandrasekhar limit.
    The explosion happens at the same stage of the star evolution in almost equal environments, which is why the mechanism is expected to be the same, with characteristic spectral properties and the same luminosity curve. 
    This type of an astronomical object, with the same luminosity (or the peak luminosity) is called a {\it standard candle}.
    Observing the standard candles at different redshifts allows us to measure the distance modulus, i.e. the difference between the apparent magnitude and the constant absolute magnitude, as the function of redshift (see \Cref{eq:mu}).
    The distance modulus is directly related to the value of luminosity distance at that redshift, which in turn depends on the expansion rate of the Universe.
    The measured value of the SN Ia flux needs to be rescaled to the B-band in the SN rest frame.
    The relation to cosmology is simply expressed via the distance modulus $\mu$:
    \begin{equation}
      \mu\equiv m_B-M_B = 5\log_{10}d_H + 5\log_{10}D_L({\mathscr C}) +25
      \label{eq:mu}
    \end{equation}
    Here $m_B$ and $M_B$ are the true apparent and the true absolute magnitudes in B-band of the SN rest frame, while $D_L$ is a \textit{Hubble constant-free} dimensionless luminosity distance that depends only on the cosmological model $\mathscr C$, but not on $H_0$ (c.f. \Cref{ssec:FLRW}).
    Having a constant $M_B$, standard candles allow us to directly relate the measured apparent magnitude $m_B$ with the cosmology.
    Although the physical ignition mechanism for the thermonuclear explosion is expected to be the same, SN Ia are not perfect standard candles, as their peak luminosities were found to vary by few magnitudes \citep{Phillips93}.
    Luckily, from the earliest SN Ia observations, various correlations between the peak luminosity and other physical properties of SN have been found.
    The most important one is the correlation between SN peak luminosity and the initial decline rate of the lightcurve \citep{Pskovskii77}.
    This correlation later led to the Phillips relation \citep{Phillips93} and finally to the absolute magnitude correction described by the lightcurve stretch factor \citep{Goldhaber01}, which was later implemented in the SALT lightcurve fitter (see \citet{Guy05,Guy07}).
    In addition to this, the JLA analysis uses a magnitude correction based on the SN colour (B-V), as proposed by \citet{Tripp98}, and on the host galaxy mass \citep{Kelly10,Johansson13,Kim14}.
    Nevertheless, there are other proposals to exploit, for example, the correlation of the SN luminosity with its metallicity \citep{Hayden13,Uemura15,Moreno16}.
    Therefore, every SN Ia can be characterised by its redshift ($z$), its apparent B-band magnitude at the peak ($m_B$), its stretch and colour factors ($s$ and $c$), and the host galaxy mass (given as $\log_{10} M/M_\odot$).
    The peak B-band absolute magnitude of SN Ia ($M_B$) is then corrected in the following way:
    \begin{equation}
      M_B^{\rm corr}=M_B+\alpha\,s-\beta\,c - \Delta M*\theta(\log_{10} M/M_\odot-10),
      \label{eq:Mcorr}
    \end{equation}
    where $\alpha$, $\beta$ and $\Delta M$ are the correction coefficients for stretch, colour and the host galaxy mass, respectively.
    These coefficients, together with $M_B^{\rm corr}$, are the nuisance parameters to be considered in the statistical analysis.
    Unlike the linear corrections for the stretch and colour, the step function $\theta$ adds a constant correction $\Delta M$ to the SN located in host galaxies with masses higher than $10^{10}M_\odot$.
    The systematic difficulty with these corrections is that they are valid only for the true values of stretch and colour, while their estimates have significant error bars.
    On average, the largest correction term is due to colour that is accounting for the reddening and for the intrinsic SN Ia colour.
    Several alternative SN Ia analysis are addressing this issue by performing the standardisation in different ways, as will be explained in the following.
    
    Besides the rest frame B-band, the observations in the NIR-band of the SN rest frame and the corresponding lightcurve can be used with the analogous NIR distance modulus \citep{Dhawan17,Weyant17}.
    In fact, the scatter of the NIR peak luminosities is about two times smaller than for the B-band, and it can be further reduced with the respective standardisation techniques for NIR lightcurve stretch \citep{Kattner12}.
    While a complete understanding of the physics of SN Ia explosion is not yet available (see also \citet{Hillebrandt00,Wang12}), the empirical standardisation relations represent the first order corrections to the peak absolute magnitude that significantly decrease the intrinsic scatter of the distance moduli, making SN Ia \textit{standardisable candles}.

    The novelty of JLA is that it analyses together all observational data of SN Ia that include light-curves and spectra, by applying the same fitting model to each supernova \citep{Guy05,Guy07}, and finally providing their astrophysical properties such as peak apparent magnitude, stretch and colour.
    Ergo, the statistical analysis of B14 has already been adopted as the standard for testing cosmological models.
    
    \subsection{Hubble residual}
    Various astrophysical and observational aspects of SN Ia are oversimplified in the JLA procedure, leading to several criticisms, proposed extensions and alternative statistical methods for SN Ia analysis \citep{Kim11,March11,Shafieloo12,Rubin15,Nielsen15,Shariff15,Li16}.
    As previously mentioned, the standardisation reduces the intrinsic scatter of absolute magnitudes, but what if a significant residual intrinsic variation remains even after the magnitude corrections?
    The remaining intrinsic differences of SN Ia progenitors, circumstellar dust, or viewing angle \citep{Goobar08,Kasen06}, which are not included in correction method, could be the origin of residual scatter.
    In addition, the unconsidered systematic errors sum up with the residuals of the standardisation technique in a way that makes them difficult to distinguish from each other.
    Wrong estimates of stretch and colour lead to wrong corrections, also contributing to the intrinsic scatter of absolute magnitude.
    \Cref{eq:mu} shows a clear degeneracy between $M_B$ and $d_H$. 
    Hence, the total scatter of corrected absolute magnitudes can be presented as the scatter in the estimates of Hubble constant.
    In fact, this remaining scatter is also called \textit{Hubble residual}.
    
    Large intrinsic scatter, $\sigma^{\rm int}$, can affect and bias the results of the SN Ia analysis and the constraints on cosmological models.
    Therefore, there have been developed many techniques for treating it.
    Commonly, $\sigma^{\rm int}$ is determined for each SN sample by adding its value in square to the apparent magnitude error bars (see e.g. \citet{Kowalski08,Conley11,Blondin11}).
    A perfect value of the added error is estimated such that it rescales the $\chi^2$ per degree of freedom to be equal to 1 for that SN sample.
    However, this method requires a fiducial cosmological fit for calculation of $\chi^2$ in each sample, or an iterative convergence fit.
    B14 use a method of restricted log-likelihood that splits the data in redshift bins and fits them separately without assumption of cosmological model inside the bin.
    The limit of their method is the statistical precision - only 7 bins are used in total, with average width of $\Delta z\lesssim0.3$.
    Certainly, increasing the apparent magnitude error bars is seemingly removing any variation, but there have been other methods proposed for dealing with the Hubble residual.
    \citet{Kim11} introduce the approach of fitting $\sigma^{\rm int}$ simultaneously with the cosmological parameters.
    In contrast to this, \citet{Shafieloo12} claims that their Bayesian method of crossing statistics can naturally take care of the intrinsic dispersion inside of the absolute magnitude posterior distribution.
    Another Bayesian hierarchical model \citep{March11} directly assumes that corrected SN Ia absolute magnitudes have values from the Normal distribution ${\cal N}(M_B^{\rm corr},\,\sigma^{\rm int})$.
    The same assumption is followed by \citet{Nielsen15} (hereafter N15), but in a frequentist likelihood approach.
    They use a common Hubble residual for the whole SN dataset, although one does expect Hubble residual to be different from one subsample to another due to the systematical differences.
    \citet{Rubin15} go further ahead by saying that unexplained dispersion should \textit{not} be attributed to absolute magnitude alone, but instead distributed over all measured parameters.
    Contributions from magnitude, stretch and colour are considered, as well as the sample dependence on dispersion\footnote{Uncertainty in the measurement of SN redshift can easily be included, as will be necessary for future data.}.
    Nevertheless, we can only hope that these residuals are purely random and do not depend on the redshift.
    For example, a possible redshift dependence of $\alpha$ and $\beta$ correction factors is often tested for \citep{Kessler09,Marriner11,Rubin15}.
    The authors found no evidence for a redshift evolution of the stretch correction factor, $\alpha$, while a decreasing colour correction factor, $\beta(z)$, was preferred in two studies of \citet{Shariff15,Li16}.
    \begin{figure}[h]
      \includegraphics[width=.48\textwidth]{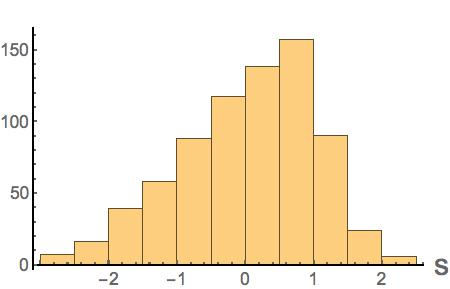}
      \hspace{.04\textwidth}
      \includegraphics[width=.48\textwidth]{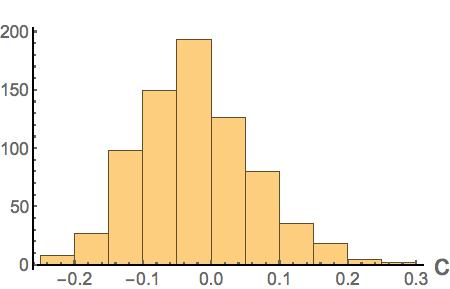}
      \caption{Histograms of stretch ($s$) and colour ($c$) factors obtained using the SALT2 fitter for all the SN in the JLA dataset.}
      \label{fig:hist}
    \end{figure}

    The correction \Cref{eq:Mcorr} is assumed to be valid only for true values of SN parameters.
    Wrong estimates of stretch and colour values undoubtedly contribute to dispersion.
    In fact, the average error bars of $s$ and $c$ are of the order of $40\%-50\%$ of their respective standard deviations.
    Hence, the face-values derived by the SALT2 fitter (and used directly in JLA approach) are not good estimates for the true values of SN parameters.
    This motivated \citet{March11,Nielsen15,Shariff15} to consider intrinsic probability distributions for magnitudes, stretches and colours.
    These authors also argue that host galaxy mass correction term (which is used in B14) has little effect on the cosmological parameters, regardless the positive tests for its existence (see also \citet{Kim14}).
    For practical convenience and mathematical simplifications distributions of $M_B$, $s$, and $c$ are taken to be Gaussian.
    Looking at histograms of intrinsic values for stretch and colour of all SN in JLA dataset in \Cref{fig:hist} (also shown in N15), assumption of normal distribution seems acceptable at first order.
    The same approach was adopted in our works \citep{Lukovic16,Haridasu17a}.
    On the contrary, \citet{Dai16} use full Bayesian approach with MCMC sampling of SN Ia lightcurve parameters to improve correction technique and minimise the effect of wrong stretch and colour estimates.
    
    A complete statistical analysis must also take care of misinterpreted SN Ia, or outliers in the sample.
    Commonly, the steps are the following: estimating intrinsic dispersion, selecting outliers, estimating coefficients of the standardisation relation ($\alpha$ and $\beta$) and finally fitting the cosmological parameters.
    Unfortunately this is an idealistic scenario, as modifying any of the steps may affect the others.
    Hence, a full analysis that performs all steps together is always preferred.
    For instance, the Bayesian method of \citet{Dai16} also suggests that 11 more outliers should be removed from JLA dataset.
    \subsection{Selection bias}
    The observed distribution of SN Ia magnitudes, but also of other intrinsic SN Ia parameters correlated with it, are shaped by the instrumentation tendency to miss observing the faintest objects, known as \textit{selection bias} \citep{Malmquist22}.
    Specifically, the supernovae that have higher $s$ and lower $c$ are more luminous and have higher probability of being observed compared to their companions on the same redshift (cf. \Cref{eq:Mcorr}).
    The same holds for $M_B^{\rm corr}$, which, as we said, has a residual variance of $\sigma^{\rm int}$.
    
\begin{figure}[h]
      \includegraphics[width=.48\textwidth]{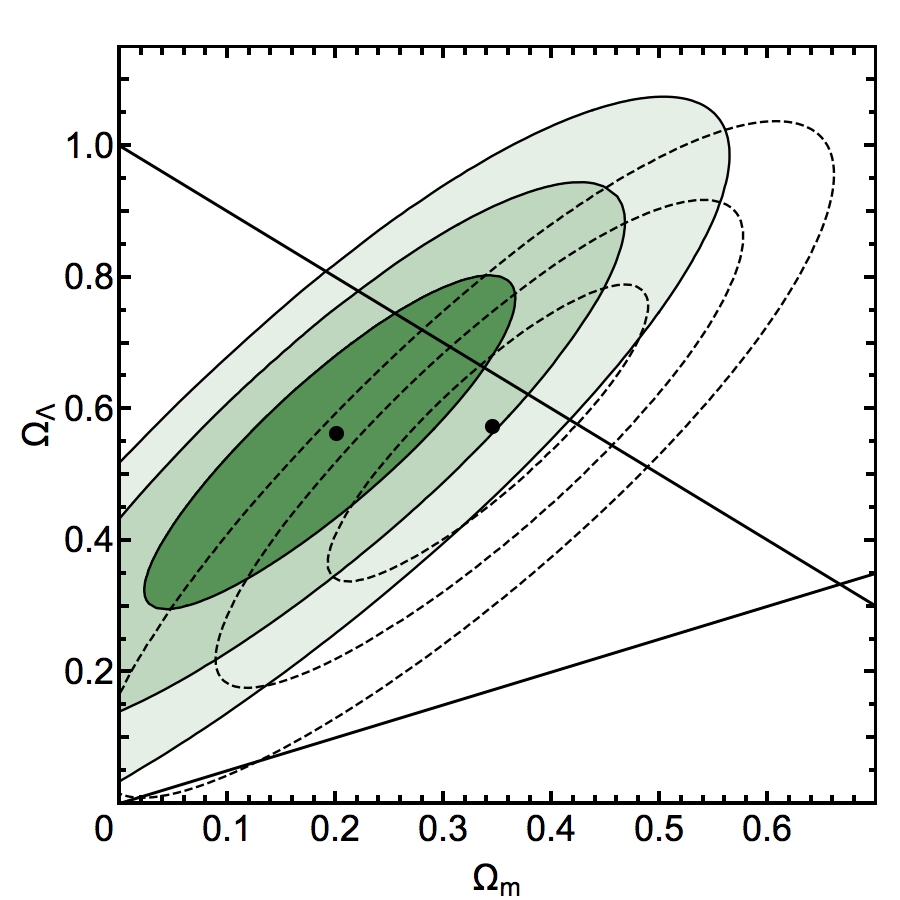}
      \caption{The shift of confidence regions from B14 analysis (green) to N15 analysis (dashed) is attributed to a difference in treatment of SN stretches and colours. The two solid lines represent the flat model and the non-acceleration model. In fact, the evidence for acceleration drops from $3.3\sigma$ for B14 to $2.9\sigma$ for N15.}
      \label{fig:BvsS}
    \end{figure}

    The effect becomes significant when telescope flux limit is reached, hence, at the highest observable redshifts of each survey.
    Although JLA includes 4 different subsamples in redshift range $0<z<1.2$, Malmquist bias is skewing the total parameter distributions seen in \Cref{fig:hist}.
    A straightforward conclusion is that \citet{March11} assumptions of Gaussian distributions for SN parameters may afflict with the final inferences of the SN analysis (see also \citet{WoodVasey07}).
    Indeed, even when used in different statistical approaches, it has an important impact on the constraints of several cosmological models \citep{Nielsen15,Shariff15,Lukovic16}.
    Namely, the evidence for acceleration coming from JLA dataset, analysed with the $k\Lambda$CDM model, is found to be less than $3\sigma$, see \Cref{fig:BvsS}.
    This initiated a range of criticism, among which, e.g., \citet{Rubin16}, who claim $4.2\sigma$ evidence using their method.
    Moreover, by adding other low-redshift observables, the evidence for acceleration is found to be strong \citep{Haridasu17,Haridasu17a}.
    
    Commonly, the selection effects are treated before the data is reduced, by usage of survey simulations and ad-hoc redshift-dependent adjustment of the observed apparent magnitudes to approximately cancel the estimated bias \citep{Conley11,Betoule14}.
    Unfortunately, the drift of stretch and colour distributions due to selection bias is not corrected.
    This remaining Malmquist bias is the origin of the shift in N15 confidence contours, shown in \Cref{fig:BvsS}.
    While the N15 Gaussian model for SN intrinsic parameters is not perfect, it is another step towards understanding SN Ia standardisation.
    
    As we explained before, the observed mean of stretch (colour) is expected to increase (decrease) with respect to redshift.
    While the explicit functional form is unknown, it is certainly different from one to another subsample.
    Assumption of sample-dependent and redshift-dependent distributions for SN parameters could be used to confront the selection bias, as done in \citet{Hinton17}. 
    \citet{Rubin16} try to model the truncation of the data ascribable to flux limit of the survey.
    Their approach is, on the other hand, criticised for using ad-hoc methods and redshift-dependent functions that include additional fitted parameters, but are not well motivated prior to application \citep{Kessler17}.
    
  \section{Baryon acoustic oscillations}
  \label{sec:BAO}
    Progressive improvements in the large-scale structure surveys like SDSS \citep{Eisenstein05} over the last decade have provided us with an extremely useful way of detecting the BAO peak in the galaxy clustering correlation function.
    The utility of the BAO peak as a standard ruler has in turn made these observations a good complementary dataset to the SN Ia observations, which were the standalone dataset to constrain cosmological models at low-redshifts.
    Until more recently \citep{Alam17} the BAO data have been presented for the observable $D_V(z)=d_V(z)/r_d$\citep{Eisenstein05}, owing to the lack of sufficient statistics to distinctly measure $d_M(z)/r_d$ and $H(z)r_d$.
    The reported $D_V(z)$ data have by far been used to constrain cosmological parameters, providing good agreement with the SN Ia data.
    Needless to say the SN Ia and BAO data together provide stringent constraints on cosmological models \citep{Lukovic16, Zhao17a, Wen17, Aubourg15, Addison17, Joudaki17} that are comparable with the high-redshift CMB constraints.
    
    In \citet{Alam17}, the results have also been presented in terms of $D_{V}$ and $AP$ (hereafter $D_{V}\&AP$) parameter space.
    Also, the covariance between the $D_{V}$ and $AP$ points is shown to be negligible.
    The $D_{V}$ observable corresponds to a spherical volume-averaged angular diameter distance, while the $AP$ variable represents an anisotropy parameter.
    They are strictly isotropic and anisotropic components of the data and they provide similar sensitivity to the cosmological parameter constraints. 
    On the contrary, $D_{M}(z)$ and $H(z)r_d$ variables (hereafter $D_{M}\&H$) are both anisotropic and the total information from the data is not necessarily distributed with the comparable sensitivity among these two components.
    Nevertheless, the $D_{V}\&AP$ can be derived from the  $D_{M}\&H$ measurements and vice versa (c.f. \citet{Alam17}).
    In a recent work \citep{Haridasu17a} we have utilised the measurements of $D_{V}\&AP$ and performed a selective analysis using the different constraints that one can obtain using these measurements individually.
    We have shown that $D_V$ and $AP$ measurements are very different in nature to constrain the cosmological parameters and can in turn be used for model selection.
    The standard approach for assessing model selection using two different data obtained from independent sources, suffers from the disadvantage that the they have different systematics.
    This gives rise to a two way explanation for the possible different results coming from the independent datasets.
    The discrepancy in the constraints obtained from fitting the model to each of the two datasets might arise due to different systematics in different data or due to the wrongly assumed fitting model (that is not the true underlying model).
    In the case of $D_{V}\&AP$ components of the BAO data, this problem is surpassed since the two components have considerably different ability to constrain the model, but they are coming from the same observation.
    A discrepancy in the constrains obtained from each of these components clearly indicates that the assumed model is incorrect (see also \citet{Verde13}).
    
    One last subtlety of the BAO data remains.
    The estimation of the sound horizon $r_d$ at drag epoch $z_d$ can be performed by either solving the Boltzmann code numerically or by using the approximate formulae that have been provided in the literature.
    This approach also provides a method to estimate the value of $H_0$ from the BAO data alone (see e.g. \citet{Wang17a}).
    The estimation of $r_d$ has for long been based on the approximate fitting formula developed in \citet{Eisenstein98a}.
    More recently, in \citet{Aubourg15} two new formulae have been proposed by also taking into account the contribution of neutrinos and possible additional relativistic species.
    These approximate formulae have been shown to be accurate up to a sub-percent level.

  \section{Cosmic expansion rate}
  \label{sec:CC}
    Along with the cosmological constant problem, another very prominently discussed issue in the standard model of cosmology today regards the value of the present expansion rate, $H_0$.
    Due to large systematic uncertainties, derived values of $H_0$ were for decades in the range $50-100$, giving rise to one of the greatest debates in astronomy - the one that still remains.
    For a review of the earlier developments in the field see \citet{Rowan85,Huchra92,Kirschner03,Jackson07,Freedman10}.
    The recent improvements are owing to the usage of space facilities, better control of systematics, and the development of different calibration techniques for distance indicators.

    The Hubble constant is an important cosmological parameter whose independent and accurate estimate is crucial for the study of dark energy dynamics, spatial curvature, neutrino physics, general relativity, linear perturbation theory, SN physics, etc.
    Therefore, the present goal is to lower the uncertainty of $H_0$ to few percents.
    While the requirements for this goal are well understood in principle and simple to list, they are extremely difficult to fulfil in practice.
    In the following subsections we revise different methods in use for estimating the value of cosmic expansion rate.
    
    \subsection{Direct measurements}
    The direct measurements of the Hubble constant fully depend on the estimates of astronomical distances, since the high precision estimates of recession velocities are easily available from the spectra.
    Various methods for measuring distances include the use of Cepheids, SN Ia, masers, tip of the red giant branch (TRGB), Tully-Fisher relation, giant HII regions, pulsar binaries, etc.
    This was also one of the Hubble space telescope (HST) key projects, which in the final result achieved a 10\% accuracy by combining several distance indicators: $H_0= 72 \pm 8 $ \citep{Freedman01}.
    The accuracy improved further in the follow-up Carnegie Hubble program reaching a 2.8\% systematic uncertainty and $H_0= 74.3 \pm 2.1$ \citep{Freedman12}.

    In order to directly probe the rate of cosmic expansion it is necessary to accurately measure distance of astronomical objects far out in the Hubble flow where their peculiar velocities are little w.r.t. expansion rate.
    SN Ia are perfect for this task as they are the best distance indicators visible at very large distances, that also provide the tightest Hubble diagram in the range $0.02\lesssim z\lesssim 0.1$.
    As we explained in \Cref{sec:SN}, SN Ia are not directly sensitive to Hubble speed due to the degeneracy with their luminosities, i.e. the intercept of the SN Hubble diagram is the sum of their mean absolute magnitude and the Hubble constant term.
    However, if one were able to assess the luminosity of SN Ia by \textit{other means}, measurements of their distance moduli on the Hubble diagram would provide the $H_0$ estimate in the aftermath (cf. \Cref{eq:mu}).

    The necessary calibration of SN Ia luminosity can be done locally with an intermediate distance indicator such as Cepheid variable stars.
    Cepheids are also one of the best distance indicators in astronomy, thanks to the relation connecting their average luminosity to the period of luminosity variation and their intrinsic metallicity.
    Once the observations become sensitive enough to probe very distant Cepheids ($\gtrsim60Mpc$ or $z\gtrsim0.02$), where their redshifts are dominated by the Hubble velocity, it will be possible to use them directly on the Hubble diagram.
    Estimating cosmological redshift of local Cepheids would require correcting the observed redshifts for relatively large peculiar velocities, but it requires the modelling of peculiar velocity field \citep{Willick01}.
    However, the most reliable methods for directly measuring cosmic distances (and $H_0$), which were also substantially developed in the last decade, use SN Ia in order to reach cosmic distances and the Hubble flow, by calibrate them locally with Cepheids or TRGBs.
    The approach of using Cepheids+SN Ia has been followed by S$H_0$ES team \citep{Riess09,Riess11,Riess16,Riess18}.
    One of the first applications by \citet{Riess05} used 2 well-observed SN Ia with typical shape and low reddening, and calibrated them with Cepheid populations observed in their host-galaxies.
    By comparing these local SN Ia to a sample of SN Ia observed farther in the Hubble flow, they found $H_0= 73\pm4\rm{ (stat) }\pm5$ (syst).
    The uncertainty was later reduced to 4.8\% with more available data \citep{Riess09}.
    
\begin{figure}[h]
      \hspace{-.5em}\includegraphics[width=0.495\textwidth,height=22em]{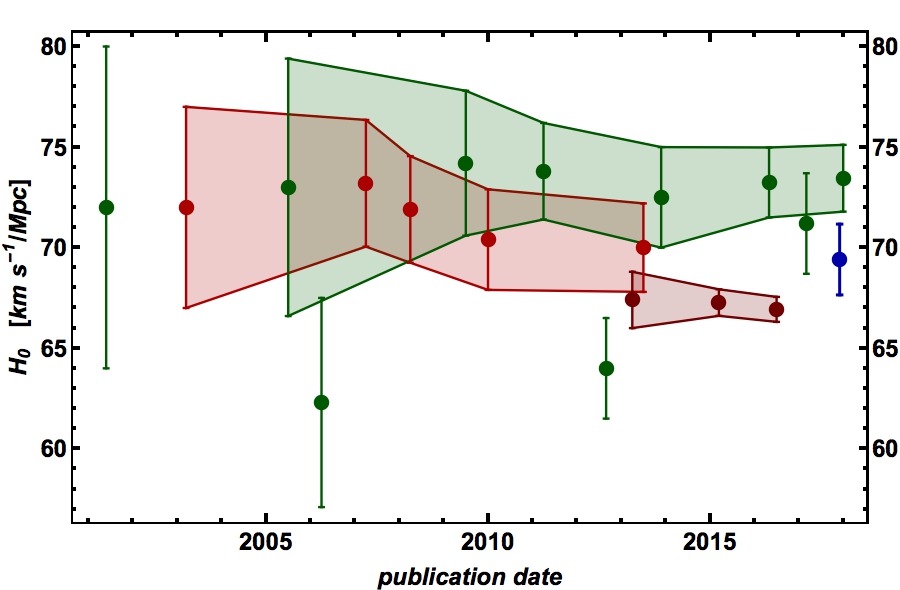}
      \caption{Historical improvement and comparison of \textcolor{green!50!black}{direct measurements} (green) vs. CMB-based estimates (\textcolor{red}{WMAP} in red and \textcolor{red!40!black}{Planck} in darker red) of the Hubble constant over the last two decades. \textcolor{blue}{Our result}, shown as the blue point, is not in significant disagreement with any of the two families. Starting point is the final result of the HST key project \citep{Freedman01}. The green shaded area gathers the results based on the Cepheids+SN Ia data from the S$H_0$ES team, in order: \citep{Riess05,Riess09,Riess11,Efstathiou14,Riess16,Riess18}. The red shaded areas are from the WMAP mission, in order: \citep{WMAP03,WMAP07,WMAP09,WMAP11,WMAP13}, and from the Planck mission, in order: \citep{Planck13,Planck15,Planck16}. We additionally show three independent direct measurements based on Cepheid and TRGB calibrations by \citep{Sandage06,Tammann13,Jang17}, in order, as non-shaded green points that have yielded lower values of $H_0$. The results are sorted by the publishing date on horizontal axis.}
      \label{fig:H0s}
    \end{figure}

    It is important to say that the zero point of Leavitt's law for Cepheids \citep{Leavitt08} must also be calibrated in order to use them as distance indicators.
    This is done using geometrical distance estimators (called {\it anchors}), such as eclipsing binaries, parallax measurements, and masers.
    A parallel to S$H_0$ES and an independent Hubble calibration programme that was using Cepheids+SN Ia provided a dissonant estimate of $H_0= 62.3\pm1.3\rm{ (stat) }\pm5.0$ (syst) \citep{Sandage06} (see \Cref{fig:H0s}).
    For the great part the mismatch between the two is caused by difference in calibration of Cepheids themselves.
    Concretely, the difference in the zero points of Cepheids' period-luminosity and period-colour relations lead to different Cepheid-based estimates for distances to SN Ia host galaxies \citep{Saha06}.
    Other quoted, possible systematic errors in this calibration of SN Ia include differences in Cepheids' populations observed locally and in SN host galaxies, such as metallicity; as well as instrumentation uncertainties between ground-based and space-based photometric zero points.
    These are summarised in \citet{Riess11} (hereafter R11), who confront them with several improvements.
    By increasing the number of Cepheids' observations and calibrated SN Ia, reducing the differences in Cepheid populations, but also by adding different independent anchors for absolute calibration of Cepheids, R11 provided a result of $H_0= 73.8\pm2.4$.
    \vspace{-1.5em}
    Several systematical uncertainties considering Cepheids treatment and estimates of anchor distances were addressed by \citet{Efstathiou14} (see also \citep{deGrijs14}), who reanalysed the data of R11 and reached the value of $H_0= 72.5\pm2.5$.
    Further improvements by \citet{Riess16} (hereafter R16) arrived to 19 selected and calibrated SN Ia, a homogeneous sample of more than 2000 Cepheid stars in SN Ia hosts, and four different anchor distances leveraging the Cepheid calibration.
    This led to a 2.4\% uncertainty of the directly estimated Hubble speed- $H_0= 73.24 \pm 1.74$.
    The result was recently corrected by \citet{Anderson17} to $H_0= 73.06 \pm 1.76$, by addressing stellar association bias for observations of Cepheids.
    Even more, the zero point calibration of Cepheids has recently been improved with parallax measurements of Galactic Cepheids, reported in \citet{Riess18} (hereafter R18) and leading to $H_0= 73.45 \pm 1.66$, as shown in \Cref{fig:H0s}.

    On the contrary, the study of \citet{Tammann13} used TRGB as independent distance indicator (instead of Cepheids) to calibrate SN Ia, reaching again a lower value of $H_0= 64.0\pm2.5$.
    However, later independent results based on TRGB-calibrated SN Ia yield somewhat higher value, such as $H_0=71.2\pm2.5$ \citep{Jang17}.
    Presently, TRGB are limited in comparison to Cepheids, regarding the number of SN Ia they cover, as well as their zero point calibration \citep{Beaton16}.
    The situation will greatly improve with the data coming from the {\it Gaia} mission \citep{Gaia16}.
    Gaia will improve geometric Cepheid calibration based on parallax measurements, it will provide a better measurement of distance to LMC, and it will also provide a parallax calibration of TRGB.
    The combination of TRGB and SN Ia will ultimately give an $H_0$ estimate completely independent of both Cepheid variables and the LMC anchor \citep{Beaton16}.
    As primary distance indicators, TRGB are more promising than Cepheids for SN Ia calibration, since they are found in galaxies of different types, while Cepheids are found only in spirals.
    
    The historically unconcordant value of $H_0$ and the complexity of measurement methods are still leaving space for suspicion considering the control of systematic errors, and consequently asking for a variety of approaches.
    A different approach towards measuring Hubble constant is followed by the Megamaser cosmology project (MCP) that is aiming to determine geometric distances to several H$_2$O megamasers in galaxies well into the Hubble flow \citep{Reid09}. 
    Their most recent estimate of $H_0= 69.3\pm4.2$ (tot) is based on four independent megamaser distances \citep{Reid13,Kuo13,Kuo15,Gao17}, but the statistical uncertainty can decrease further with more megamaser observations.
    Several reanalyses of the S$H_0$ES data were performed in the search for systematic errors, e.g. \citep{Efstathiou14,Zhang17,Feeney17,Follin17}.
    The high value of local expansion rate obtained from Cepheids+SN Ia was confirmed using observations in the NIR-band of the SN rest frame- \citet{Dhawan17} found $H_0=72.8\pm3.2$.
    A very recent estimate of $H_0= 71.0\pm2.8\rm{ (stat) }\pm2.1$ (syst) \citep{Arenas18} is based on a different technique that uses HII galaxies and giant HII regions as distance indicators.
    Their estimate of distance is emerging from the correlation between the turbulent emission lines velocity dispersion of the ionised gas and their integrated H$\beta$-line luminosity \citep{Terlevich81}.
    Another way of measuring $H_0$ emerges from the effect of \textit{strong gravitational lensing}.
    Light coming from high redshift sources bends around the clustered matter at lower redshifts, while the deflected light rays result in multiple images of the same source.
    Considering the time delay between these multiple images (TDSL) and assuming $\Lambda$CDM model, \citet{Bonvin16} get $H_0= 71.9^{+3.0}_{-2.4}$ (see also \citet{Liao17} for TDSL of gravitational waves).
    All of the cited direct $H_0$ measurements up to date are consistent among them.
    
    The latest multi-messenger detection of neutron star merger using both gravitational waves and electromagnetic waves \citep{Abbott17b,Abbott17c,Goldstein17,Savchenko17}, provided for the first time a possibility to use the gravitational wave detection as a standard siren for measuring luminosity distance to the source and, with it, the present cosmic expansion rate \citep{Abbott17n}.
    This event provided a less precise measurement of $H_0=70^{+12}_{-8}$, but it demonstrates a very promising technique.
    One more exciting method of measuring $H_0$ starts with estimation of distances to the clusters of galaxies using the ratio of their X-ray emission and Sunayev-Zel'dovich effect (SZ).
    Although the last estimate with this method from \citet{Bonamente06}, who uses 38 clusters, has very large error bars due to older data and large systematics, the latest SZ cluster measurements from Planck collaboration data are bound to provide a more competitive figure \citep{Planck15SZ,Bourdin17}.

    \subsection{$H_0$ as a cosmological parameter}
    A few methods listed above are not exactly direct measurements of Hubble speed, since they rely on the assumption of a fiducial cosmological model for the data analysis.
    Furthermore, there are many different classes of astrophysical observables that can be fitted with a cosmological model where $H_0$ constitutes as a fitting parameter.
    The strongest constraint is coming from the CMB anisotropy observations.
    In fact, the latest result by \citet{Planck16} reports $H_0=66.93\pm0.62$ based on the full CMB analysis (see their Table 8, last column).
    It is clear that this model-dependent estimate has lower value of $H_0$ compared to the most recent direct measurements.
    Specifically, there is a large tension with the measurement based on Cepheids and SN Ia- the latest one from \citet{Riess18} is $3.3\sigma$ away from P16 value (cf. \Cref{fig:H0s}).
    However, P16 is not in tension with the direct measurement of \citet{Tammann13}, who got $H_0= 64.0\pm2.5$ based on SN Ia calibrated with TRGB observations.
    The straightforward question is whether the discrepancy in $H_0$ estimates is asking for even better control on systematics in these experiments, or it points towards new physics beyond what is by now commonly known as the concordance model.

    Direct measurements of the cosmic expansion rate at distant redshifts is also possible.
    \citet{Jimenez02} proposed the differential age (DA) method, which uses pairs of passively-evolving red galaxies found at similar redshifts (let's say: $z$ and $z+\Delta z$) to estimate $dz/dt$.
    Characteristic spectral properties of these early-type, massive, elliptical galaxies help us to determine their age, which has made them popular as \textit{cosmic chronometers}.
    Assuming that the same type of galaxies at close-by redshifts formed at almost the same cosmic time, the measured difference of their ages is equal to difference in our perceived age of the Universe at these two close-by redshifts: $\Delta t(z)=t(z+\Delta z)-t(z)$.
    The Hubble parameter at redshift $z$ is easily related to this measure
    \begin{equation}
      H(z)={1\over a}{\diff a\over\diff t}=-{1\over1+z} {\diff z\over\diff t}\approx-{1\over1+z} {\Delta z\over\Delta t}
    \,.\end{equation}
    
    There have been by now 31 uncorrelated $H(z)$ points reported by different sources, all of which used the same model for galactic age estimation \citep{Simon05,Stern10,Moresco12b, Moresco16a, Moresco15, Zhang14,Ratsimbazafy17}.
    Obviously, the estimate of $H_0$ is possible to obtain from a fit of these points assuming a specific theoretical form of the $H(z)$ function.
    Since the DA method has a limited precision, the cosmological fit of CC does not have very strong constraints on fitting parameters, including $H_0$ \citep{Chen17}.
    Nevertheless, CC can be combined with other low-redshift data that strongly constrain the functional slope of the cosmic expansion, $H(z)/H_0$, leading to substantial improvements on the estimate of the scaling constant, $H_0$, from CC data.

    Often, the BAO data are used for determination of the Hubble parameter by means of the fitting formula \citep{Eisenstein98a} for the value of sound horizon at drag epoch, which depends on $H_0$ (see e.g. \citet{Gaztanaga09,Aubourg15,Cheng15,Wang17a}).
    We do \textit{not} use Eisenstein fitting formula for estimating $H_0$ from the BAO data alone, but instead use $H_0 r_d$ as the fitting parameter.
    In such a way BAO analysis alone does not give direct constraint on $H_0$ nor $r_d$, as these two parameters appear degenerate.
    In order to extend the discussion on the tension of $H_0$ values coming from direct measurements of R16 and $\Lambda$CDM-based analysis of P16, we adopted the approach of estimating $H_0$ from low-redshift observables as an independent result \citep{Lukovic16,Haridasu17a}.
    Specifically, our result for $H_0$ is arising from the CC data, while SN Ia and BAO data constrain \textit{only} the other cosmological parameters that define $H(z)/H_0$.

    \section{Constraints on cosmological models}
    \label{sec:res}
In this section we summarise the constraints obtained using the data aforementioned. We consider both N15 and B14 Supernovae analysis methods and also compare the constraints obtained from these methods also in a combined analysis. A joint analysis of SN Ia, CC and BAO datasets allows us to break the degeneracies between parameters such as $M_b$, $H_0$ and $r_d$ and to compare these constraints with those obtained using high-redshift CMB or local distance-ladder measurements of $H_0$. As the three datasets are independent, we evaluate the total likelihood as the product of the likelihoods of each individual datasets. Therefore, for FLRW models $\cal{L}_{\textrm{tot}}=\cal{L}_{\textrm{ SN}}\cal{L}_{\textrm{ CC}}\cal{L}_{\textrm{ BAO}}$, while for LTB cosmologies we do not consider the BAO dataset.

We use the well known Akaike information criteria (AIC) \citep{Akaike74} and Bayesian information critereia (BIC) \citep{Schwarz78} for model selection. Comparing the AIC or BIC values obtained for two different models using the same data, the model with lower value of AIC or BIC remains to be the preferred model (c.f. \citep{Shafer15,Lukovic16,Haridasu17,Haridasu17a}).

The results in this section are presented for the comparison of the different SN Ia analysis methods and then the comparison of different components of the BAO data and then we proceed to comment on our joint analysis using all the datasets in the framework of Friedmann cosmologies. We finally present our constraints on the inhomogeneous LTB model.

\subsection{Results focusing on SN Ia data}
Typically, SN Ia parameters, and specifically the peak absolute magnitude, are derived directly from the fit, since they depend on cosmological model.
The degeneracy of absolute magnitude and Hubble radius (cf. \Cref{eq:mu}) can be broken once SN Ia are fitted together with another observable that is directly sensitive to $H_0$.
In our combined fit this is possible thanks to the cosmic chronometers that enable us to constrain the cosmic expansion rate and the SN Ia luminosity at once.
\Cref{tab:SNintB} lists our best estimate for the mean corrected value of SN Ia absolute magnitude, but also all other SN Ia intrinsic parameters obtained in the joint $\Lambda$CDM fit of JLA+BAO+CC.
  \begin{table}[h!]
    \begin{center}
      \footnotesize
      \caption{SN Ia intrinsic parameters from the combined $\Lambda$CDM fit of JLA+BAO+CC using B14 method for SN standardisation.}
      \begin{tabular}{|c|c|c|c|}
        \hline
        $M_B^{\textrm corr}$ & $\Delta M$ & $\alpha$ & $\beta$\\
        \hline
        $-19.07 \pm 0.06$ & -0.07 $\pm$ 0.02 & 0.141 $\pm$ 0.007 & 3.10 $\pm$ 0.08 \\
        \hline
      \end{tabular}
      \label{tab:SNintB}
    \end{center}
  \end{table}
Our result for the corrected B-band absolute magnitude perfectly agrees with \citet{Richardson14} estimate $M_B^{\textrm corr}=-19.25\pm0.20$, and also with B14 estimate $M_B^{\textrm corr}=-19.05\pm0.02$.
Both of them use a fixed value for $H_0=70$ that is very close to our best fit of $H_0=69.4$, which is a necessary premise for the agreement on absolute magnitude values.
Although we use the same method of SN standardisation as in B14, our uncertainty for $M_B^{\textrm corr}$ is larger since we do not assume a fixed value for $H_0$ and, hence, the uncertainty of our $H_0$ estimate enters in the error bar of $M_B^{\textrm corr}$.

Similarly, we present the SN parameters obtained with the N15 method.
In this model SN is characterised with eight intrinsic astrophysical parameters: six for the normal distributions ${\cal N}(M_0,\sigma_{M_0})$, ${\cal N}(s_0,\sigma_{s_0})$ and ${\cal N}(c_0,\sigma_{c_0})$, and two constant correction coefficients $\alpha$ and $\beta$ of \Cref{eq:Mcorr}.
\begin{table}[h!]
  \begin{center}
    \footnotesize
    \caption{Results for the SN Ia intrinsic parameters from the combined $\Lambda$CDM fit of JLA+BAO+CC using N15 model.}
      \begin{tabular}{|c|c|c|c|}
        \hline
        $M_0$ & $\sigma_{M_0}$ & $s_0$ & $\sigma_{s_0}$\\
        \hline
        $-19.13 \pm 0.05$ & 0.107 $\pm$ 0.005 & 0.042 $\pm$ 0.038 & 0.933 $\pm$ 0.027 \\
        \hline\hline
        $c_0$ & $\sigma_{c_0}$ &$\alpha$ & $\beta$\\
        \hline
        -0.020 $\pm$ 0.005 & 0.071 $\pm$ 0.002 & 0.134 $\pm$ 0.006 & 3.059 $\pm$ 0.087 \\
        \hline
      \end{tabular}
      \label{tab:SNintS}
    \end{center}
  \end{table}
  The results in \Cref{tab:SNintS} are consistent and almost the same as in our earlier work \citep{Lukovic16}, although here we used a substantially updated BAO data in the joint fit (see \Cref{sec:BAO}).
  The N15 method specifically differs in the treatment of residual variation from the previous method, where it is added to the data error bars. We can see that the variation of the corrected absolute magnitude, $\sigma_{M_0}=0.11$, is not small compared to the typical stretch and colour correction terms, $\alpha\,\sigma_{s_0}=0.13$ and $\beta\,\sigma_{c_0}=0.22$, respectively.
It is also larger than the typical host galaxy mass correction term used in the previous method $\sim\frac12\Delta M=0.035$.

Interestingly enough, we find the best-fit results for the SN parameters to be nearly independent of the cosmological model under consideration.
As the only exception, a stronger correlation between cosmological and SN parameters is seen for the value of the mean absolute magnitude that is related to $H_0$ and through it with other cosmological parameters.
Peculiarly, the $\Omega_m$ shift in \Cref{fig:BvsS} affects also the joint analysis that uses N15 model, leading to a lower best-fit of $H_0=68.2$ and a lower mean value of the absolute magnitude, $M_0=-19.13$ than in the first model. We also note an excellent agreement of $\alpha$ and $\beta$ correction factors, even though the two analyses use different SN modelling.

  We have presented the results for the joint analysis using the N15 method of the SN Ia analysis in \citep{Haridasu17}.
  Although, it remains that the evidence for acceleration from the SN Ia data alone reduces in the $k\Lambda$CDM plane, while the joint analysis and $w$CDM model do predict a strong evidence for the same.

\begin{figure}[h]
\includegraphics[width=0.45\textwidth]{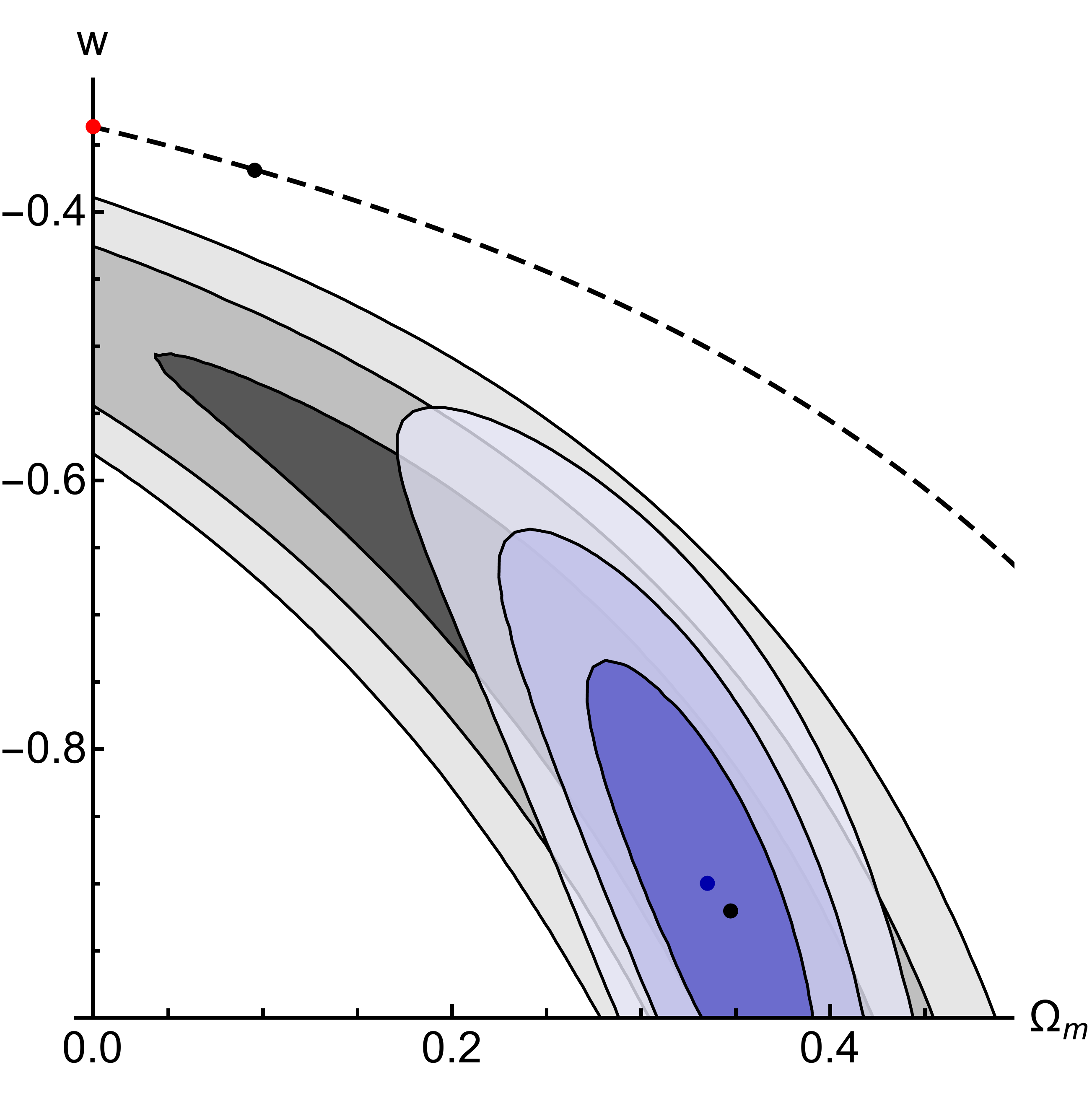}
\caption{The grey and violet confidence regions are obtained from the SN Ia alone and the joint analysis, respectively. The 1$\sigma$, 2$\sigma$ and 3$\sigma$ confidence regions in the $\Omega_{m} - w$ parameter space for the $w$CDM. The dashed curve shows the no-acceleration criterion: $w = -1/(3(1-\Omega_m))$. The black point marks the $5.38\sigma$ evidence for acceleration from joint analysis. This figure is taken from \citet{Haridasu17}.}
\label{fig:con}
\end{figure}

The red point in \Cref{fig:con} represents the power-law model with best-fit $n= 1.08 \pm 0.04$ (see \Cref{eqn:wn}). However, note that this joint analysis is performed only with the isotropic $D_V$ BAO data and also with the inclusion of gamma ray burst (GRB) dataset\footnote{Please refer to \citet{ Wei10, Amati02, Amati08, Amati09, Haridasu17} for more detailed discussion on the GRB dataset.}. The use of GRB allows us to explore the models at much higher redshifts, contributing to the necessary information criteria to perform model selection.

\begin{table}[h]
\caption{$\Delta$(AIC)  and $\Delta$(BIC) comparisons for models with $\Lambda$CDM as the reference. 'Joint' corresponds to the joint analysis with SN+BAO+CC+GRB datasets.}
\label{tab:aicold}
\begin{center}
\footnotesize
\begin{tabular}{|c|c|c|c|c|}
\hline
& $\Delta\text{(AIC)}_{\text{Joint}}$& $\Delta\text{(AIC)}_{\text{SN}}$& $\Delta\text{(BIC)}_{\text{Joint}}$& $\Delta\text{(BIC)}_{\text{SN}}$ \\
\hline
Power-law & 28.02 & 2.0 & 28.02 & 2.0 \\
$R_{h}=ct$ & 30.83  & 21.79 & 26.05 & 17.20 \\
Milne & 66.39  & 9.78 & 61.62 & 5.19 \\
\hline
\end{tabular}
\end{center}
\end{table}

\subsection{Results focusing on BAO data}
The current BAO data owing to its precise measurements now reigns to be the most constraining dataset in the redshift range of $0<z<2.5$. In \citet{Haridasu17a} we have utilised the most recent BAO datasets to test the constraints obtained using the isotropic $D_V$ and anisotropic $AP$ components separately. Interestingly, we find that the $AP$ measurements give a lower value of $\Omega_{m}$ for $\Lambda$CDM and all the standard extensions, while the $D_V$ measurements give higher values of the same with very similar power to constrain the mean value. In \Cref{tab:baofitlcdm} we show the best-fit parameters to $ \Lambda $CDM model using individual observables and their combinations. The $\Omega_m$ estimates obtained using the $D_V$ and $AP$ components separately for $\Lambda$CDM model show a mild disagreement of $2.1 \sigma$.
 \citet{Addison17} have reported a tension of $\sim$ 2.4$\sigma$ between the low-redshift galaxy-clustering BAO data and the older high-redshift Lyman-$\alpha$ data. We find this tension still remains, however slightly lowered, as the Lyman-$\alpha$ measurements have been revised with larger dispersions \citep{Bautista17,MasdesBourboux17}. However, a detailed comprehension on this tension needs to be further explored.

{\renewcommand{\arraystretch}{1.75}%
\begin{table}[h]
\begin{center}
\caption{Fit parameters to the BAO data in the four different formalisms for $\Lambda$CDM model.}
\label{tab:baofitlcdm}
\footnotesize
\vspace{0.2in}
\begin{tabular}{cccc}
\hline
\hline
 Data &$\Omega_{m}$ & $H_{0}r_{d}[\text{Km/s}] $  \\
\hline
$AP$ & $0.225^{+0.045}_{-0.040}$ &  -    \\

$D_{V}$ & $0.358^{+0.043}_{-0.038}$ & $9840^{+204}_{-212}$  \\

$D_{V}\&AP$ & $0.285^{+0.019}_{-0.017}$ & 10182 $\pm$ 139  \\

$D_{M}\&H$ &  $0.288^{+0.019}_{-0.018}$ & 10162 $\pm$ 139 \\
\hline
\end{tabular}
\end{center}
\end{table}
}

We find that the $AP$ component plays a very crucial role, as it gives necessary contributions to obtain flatness in $k\Lambda$CDM and shows better sensitivity to the EOS parameter in $w$CDM model. As we perform a comparative study of $D_V\, \textrm{vs.}\, AP$, we refrain from using the three $D_V$ only measurements at $z= 0.106, 0.15, 1.52$. However, we find that the e.g. $\Lambda$CDM constrains are only mildly affected by their inclusion (see also \citet{Addison17}).

\subsection{Joint analysis}

In this section we summarise the constraints obtained from the joint analysis of BAO+SN+CC data. In \cref{tab:joint} we show our findings from \citet{Haridasu17a} on the constraints on the cosmological parameters. We find that the estimates for $H_0$ and $\Omega_m$ are consistent among all the models considered here. Note the low value of $\Omega_m$ obtained for the $w_0w_a$CDM model alone.

\begin{figure}[h]
\includegraphics[width=0.45\textwidth]{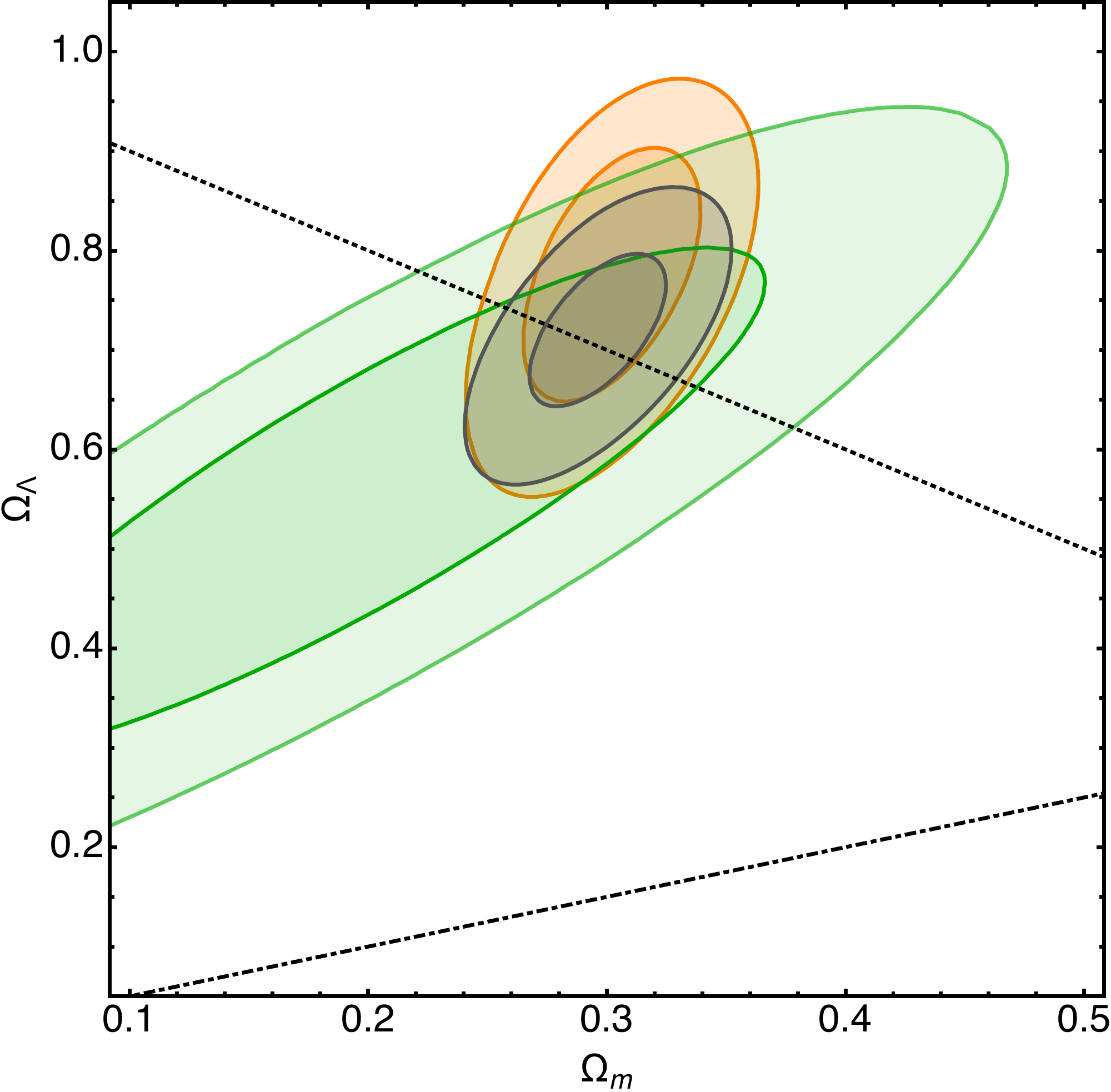}
\caption{The orange, green and black contours correspond to $D_{M}\&H$, SN Ia and combined analysis, respectively. We show here the 68\% and 95\% confidence level contours. The dotted line correspond to the flat model, the dot-dashed line marks the transition between the accelerated and non-accelerated regimes.}
\label{fig:con1}
\end{figure}

{\renewcommand{\arraystretch}{1.75}%
\begin{table}[h]
\begin{center}
\caption{Best-fit estimates and parameter constraints using the $D_M\&H$+SN+CC data.}
\label{tab:joint}
\footnotesize
\vspace{0.2in}
\resizebox{\columnwidth}{!}{%
\begin{tabular}{cccccc}
\hline
\hline
 Model &$\Omega_{m}$ &$H_{0}$ & $ w_{0}/w $ & $w_{a}$&$\Omega_{\Lambda}$   \\
\hline

$\Lambda$CDM & $0.292^{+0.016}_{-0.015}$  & $69.41 \pm 1.76$ & - & - & - \\

$k\Lambda$CDM & $0.296 \pm 0.024$  & $69.62^{+2}_{-1.98}$ & - & - & $0.722^{+0.064}_{-0.067}$ \\

$w$CDM & $0.285 \pm 0.018$  & $68.61^{+1.93}_{-1.91}$ & $-0.921^{+0.08}_{-0.081}$& - & - \\

$w_{0}w_{a}$CDM & $0.195^{+0.084}_{-0.23}$  & $68.75^{+1.95}_{-1.92}$& $-0.902^{+0.222}_{-0.125}$& $0.838^{+0.217}_{-0.655}$ & - \\

$kw$CDM &$0.311 \pm 0.025$ & $69.27^{+2.0}_{-1.97}$& $-0.828^{+0.075}_{-0.089}$& - &$0.872 \pm 0.107 $ \\

\hline
\end{tabular}%
}
\end{center}
\end{table}
}

Also, $D_M\&H$ data alone shows that the standard model is in agreement only at $2.2\sigma$ in $kw$CDM model. Here the correlation between the two free parameters $\Omega_k$ and $w$ is such that, reducing either of the degree of freedom tends to converge towards $\Lambda$CDM. We have shown that assuming the A15 formula while fitting the $w_0w_a$CDM model tends to improve the agreement with the standard model. While not using the A15 standard model is agreeable at $1\sigma$. As it was shown that the A15 formula is to sub-percent level accurate, it remains to be understood if this result indicates a necessary modification to the physics of the early universe. Among the models we have tested, the one parameter extension models - $k\Lambda$CDM and $w$CDM - are very well in agreement with the standard model. The two parameter extension models - $kw$CDM and $w_0w_a$CDM - clearly need to be better understood, and are  possibly indicating a need for new physics (see also \citep{Ooba17,Park18}).

\subsection{Comment on acceleration}
The BAO data have been able to provide much more significant evidence for the same in $k\Lambda$CDM scenario at $5.8 \sigma$ (see also \citep{Ata17}). However, in the $wCDM$ scenario BAO data has been unable to provide a significant evidence \citep{Haridasu17a, Lonappan17}. Finally, in the joint analysis we find an evidence of $8.4 \sigma$.
Non-accelerating power-law cosmologies have also been of keen interest in this respect. \citet{Tutusaus17} have implemented several evolving magnitude models in the SN Ia analysis and showed that in such scenarios the evidence for acceleration significantly reduces and it yet remains to be understand the true nature of the SN Ia physics/phenomenon.

The high-redshift CMB data - even with its very strong ability to constrain the cosmological models - is only able to indicate an accelerated phenomenon for a constant EOS dark energy scenario. CMB data by itself is unable to provide any significant constrains or indications for a dynamical nature of dark energy. In \citep{Zhao17a}, a $3.5 \sigma$ evidence for a dynamic dark energy has been quoted by performing an hybrid analysis to reconstruct the EOS using model-independent Principal component analysis with a model- dependent Friedmann background.

    \subsection{Constraints on LTB model}
    Besides the classical FLRW cosmologies we have explored an alternative model of inhomogeneous Universe.
    Ever since the discovery of accelerated expansion, based on SN Ia data, LTB models have been widely tested against cosmological observables \citep{Celerier00,Alnes06a,Sundell15,Zhang15}.
    In particular the analyses of void models using SN Ia data indicated that we live in a large underdensity of the Gpc scale, e.g. \citet{GBH08}.
    Several matter density profiles have been considered, with the general conclusion that the size of the void is more robustly constrained, while the assumption about the shape is flexible \citep{February10}.
    
    Linear scalar perturbation theory for background LTB model was needed to be developed from the scratch as it is different from FLRW scenario \citep{Zibin08,Clarkson09,Zibin11}.
    Likewise, \citet{Alonso10} address several aspects of numerical simulations of large scale structure evolution in an inhomogeneous background.
    Not surprisingly, for analysis of the BAO data in inhomogeneous models, one has to start from the first principles.
    The distance measures and the proper physical length depend not only on time but also on the radial coordinates, demanding an adequate treatment of the BAO peak observations.
    \citet{Zumalacarregui12} use BAO measurements to rule out giant void models, independent of other observational constraints.
    \citet{Vargas17} found that BAO constraints of LTB model disagree from those coming from SN Ia. 
    \citet{Bull12} showed that giant void models with inhomogeneous Big Bang times can be constructed to fit the SN Ia data, WMAP small-angle CMB power spectrum, and the local measurements of $H_0$ simultaneously.
    However, \citet{Amendola13} show that the predicted kinematic Sunyaev-Zeldovich signal in such models is severely incompatible with existing constraints, which together with other cosmological observables rules out the giant void models.
    \begin{figure}[h!]
      \includegraphics[width=\columnwidth]{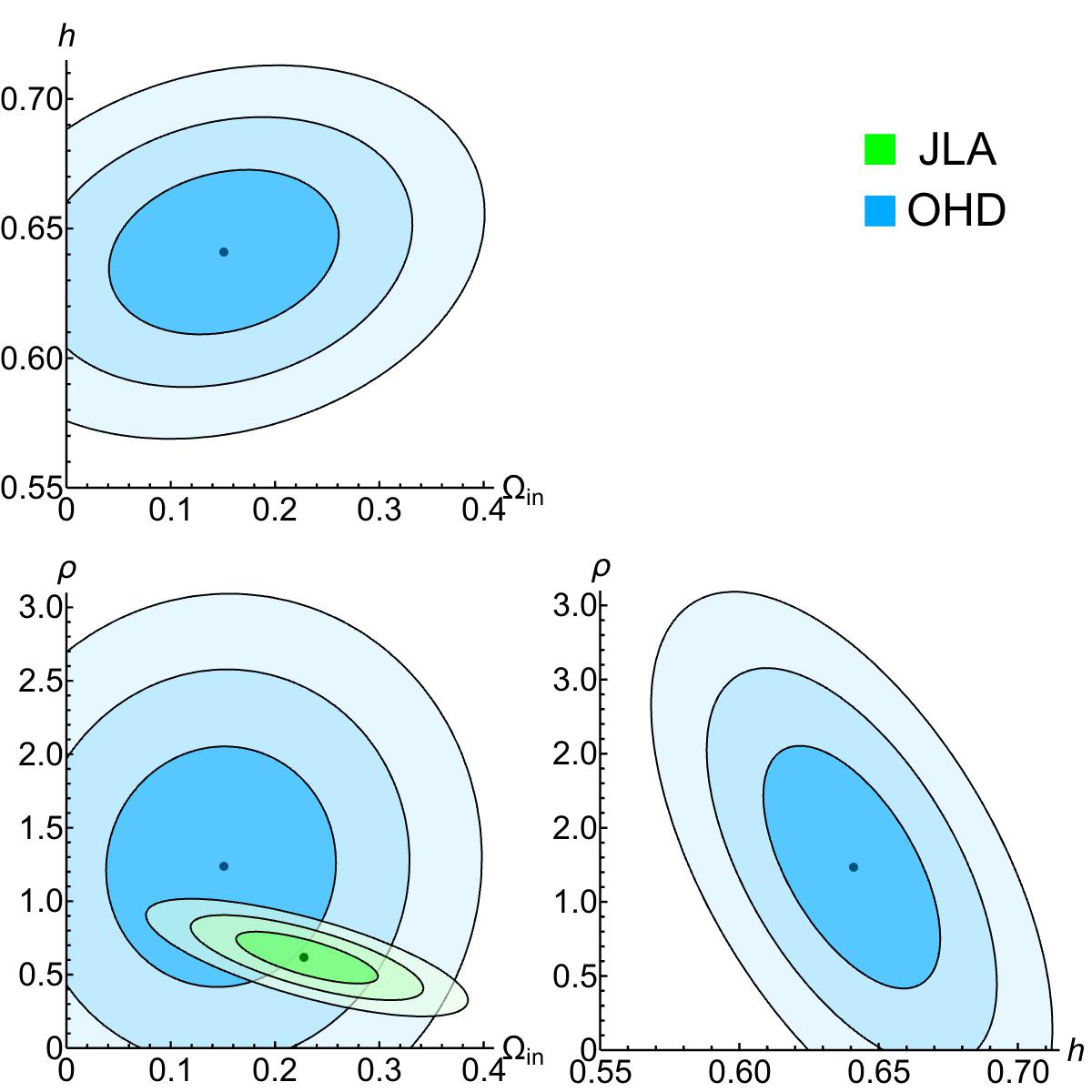}
      \caption{$1\sigma$, $2\sigma$ and $3\sigma$ confidence regions for the LTB model to the single datasets as indicated in the top right panel. This figure is taken from \citet{Lukovic16}.}
      \label{fig:LTB_ell}
    \end{figure}

    In \citet{Lukovic16} we have tested LTB model against SN Ia and CC data, but we did not use BAO data in order to avoid some subtle issues regarding the effects from the spatial dependence of the background dynamics.
    The parameter constraints are shown in \Cref{fig:LTB_ell}.
    Consistently with previous studies, our results show that the void model is capable of explaining the analysed data without DE component, but by breaking the Hubble scale homogeneity.
    The large size of the void brings questions regarding our spatial position inside of it.
    \citet{Alnes06b} found that the observer has to be located almost at the centre of the void, $\sim1\%$ of the void radius, in order not to introduce a too large and stay consistent with the dipole measurements coming from the CMB spectra.
    Constraining the observer to be effectively at the on-centre location is a very special and improbable position that is against the Copernican principle.
    
    Besides the void size, we estimated the value of $H_0=64.2\pm1.9$, which is in strong tension with the direct measurements, and even lower than the $\Lambda$CDM-based estimate.
    Indeed, the classical void model without cosmological constant is known to have a lower fitting value of $H_0$ \citep{Nadathur11}, which was also used as an argument for dismissing inhomogeneous cosmologies \citep{Riess11}.
    The peculiarity of LTB model is that value of the present expansion rate is not the same inside the void and for the background solution.
    The low matter density at the void centre is boosting the expansion, and it ensures the higher value of present expansion rate inside than outside the inhomogeneity.
    
    The incompatibility of LTB models with cosmological observations, together with characteristic radially dependent expansion rate, encouraged remodelling the inhomogeneous cosmologies \citep{Moffat16}.
    A cosmological model that features a large scale matter inhomogeneity in a background $\Lambda$CDM model is performing much better in data analysis than the classical LTB model without cosmological constant \citep{Hoscheit17}.
    \citet{Tokutake17} demonstrate how this $\Lambda$LTB model is able to provide a better agreement with local Hubble rate measurements.
    Even more, its radial profile of the expansion rate allows for a lower value of the asymptotic background solution, resolving the discrepancy between the locally measured and the CMB-estimated values of $H_0$.
    
    We have considered a flat $\Lambda$LTB solution that has a matter density profile followed by an inhomogeneous DE density profile in a flat geometry.
    Our analysis of low-redshift data showed that flat $\Lambda$LTB is disfavoured w.r.t standard $\Lambda$CDM model \citep{Lukovic16}.
    However, we note that the situation changes for an open curvature $\Lambda$LTB model that has both matter density and curvature profiles, as used in \citet{Tokutake17}.
    The cosmic metric that is based on $\Lambda$LTB solution locally and asymptotically converges to $\Lambda$CDM metric provides a framework for studying the effects of inhomogeneities in standard cosmological model.
    Furthermore, it can play an important role in precise reconstruction of the cosmic metric that is necessary for future data \citep{Vallejo17}.
    
    \section{Hubble constant issue}
    \label{sec:H0}
        Ever since the HST key project (KP hereafter), the extraordinary improvements in the direct determination of the distance ladder were followed by the strikingly stringent estimates of Hubble constant coming from CMB anisotropy observations.
    The most recent direct measurement for the present expansion rate of $H_0= 73.45 \pm 1.66$ by R18 has a large discrepancy of $3.7\sigma$ compared to the latest estimate coming from P16, which is based on the concordance model, $H_0= 66.93 \pm 0.62$.
    This situation gave rise to many recent discussions \citep{Freedman17}.
    The first obvious possibility for the origin of tension may be the presence of significant systematic errors in one or both estimates.
    Otherwise, taking the two results at face values indicates that $\Lambda$CDM model is unable to concord with the description of the early universe with the local measurements.
    One suggested modification of the standard model in order to arrive to higher value of $H_0$ from the CMB analysis is to consider larger number of relativistic species in the early universe \citep{Bernal16}.
    Indeed, increasing the number of neutrino species in $\Lambda$CDM model reduces the size of the sound horizon, which in effect changes the fit to CMB data.

\begin{figure}[h!]
      \includegraphics[width=\columnwidth]{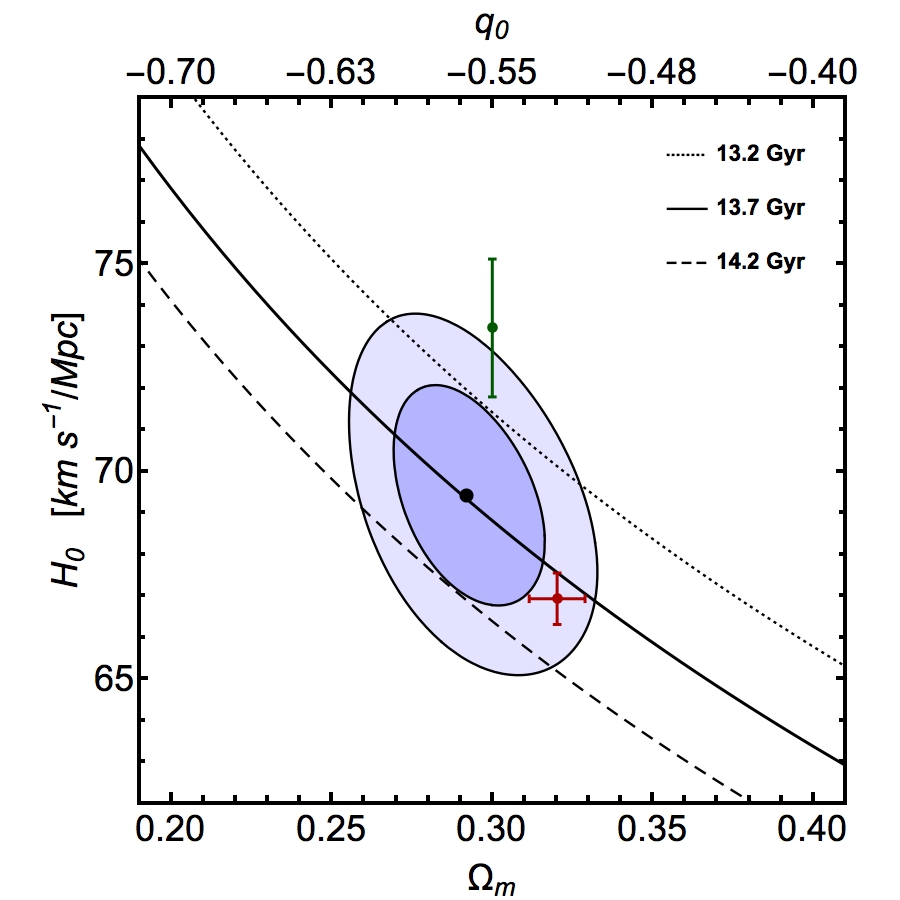}
      \caption{Theoretical iso-ages for the $\Lambda$CDM model, corresponding to ages of ($13.7\pm0.5$) Gyr are shown in the $\Omega_m-H_0$ plane, together with the 1$\sigma$ and 2$\sigma$ confidence regions resulting from the joint JLA+BAO+CC analysis for $\Lambda$CDM model. The red cross and the green line are the results of P16 and R17, respectively.}
      \label{fig:isoch}
    \end{figure}

    The best way to check whether the origin of the tension lays in systematical errors or in wrong theoretical predictions is to aim for other independent results that confirm one of the two.
    Most of the direct measurements prefer high value of $H_0$, but none of them is reporting so small uncertainty as R18 and are not in significant tension with P16 (cf. \Cref{fig:H0s}).
    Our approach was to use independent low-redshift ($z\lesssim2$) data and provide the estimate of the expansion rate that is based on the observations of cosmic chronometers.
    In order to fit the data we used a number of different FLRW and inhomogeneous cosmologies, so our result depends on model assumptions and the interplay of cosmological parameters.
    Nevertheless, our conclusions about the standard $\Lambda$CDM model are fully representable in $\Omega_m-H_0$ plane.
    Interestingly, our result of $H_0=69.41\pm1.76$ lays in the middle of the two main values in tension, and is not in a significant disagreement with any of them (see \Cref{fig:isoch}).
    In fact, it is $\sim 1.3 \sigma$ away from the estimate of P16 and $\sim 1.6 \sigma$ away from the direct measurement of R18.
    The result of S$H_0$ES team emerges from the fit of SN Ia at $z\lesssim0.15$, which are calibrated with Cepheids.
    In order to fit the SN Ia sample, R16/R18 does not use a cosmological model but a Taylor expansion of Hubble parameter around $z\approx0$ with an assumed deceleration parameter of $q_0=0.55$ (corresponds to $\Omega_m=0.3$ in $\Lambda$CDM).
    The Hubble constant is simply the offset value of this Taylor expansion.
    As their final result depends on this assumption, R16 add the systematical uncertainty of $H_0$ estimate due to small variation of $q_0$ parameter.
    Although our result is between the two, it is still decreasing the probability for the concordance model, since we find using the {\it index of inconsistency} measure (see \citet{Lin17}) that the total disagreement of all three is 4.2 \citep{Haridasu17}.
    
    Searching for a minimal variant of standard cosmological model that provides better description of the astrophysical data, one has to consider different modifications that are correlated with estimate of $H_0$ parameter.
    Broadly speaking, the extended CMB analyses that consider more free parameters than the base $\Lambda$CDM model are expected to have more loose constraints on present expansion rate and decrease the tension.
    However, when the CMB data is combined with direct expansion rate measurements, the constraints on other cosmological parameters are bound to change from their base values in order to accommodate for the $H_0$ shift \citep{Riess16}.
    This is exactly the case when we relax the assumption on the neutrino species only to conclude that a joint CMB+$H_0$ fit arrives at higher number of relativistic species than in base cosmological model.
    In any case, this change would not affect our conclusions for the $\Lambda$CDM model, since our procedure does not depend on assumptions about the sound horizon length and our $H_0$ value is determined by CC.
    Similarly, \citet{DiValentino16} considers many additional cosmological parameters in an extended $\Lambda$CDM model, and by reevaluating the CMB analysis with the R16 measurement they do not find the tension.
    However, their estimate of $w = -1.29_{-0.12}^{+0.15}$ is $\sim 2.2\sigma$ away from our result of $w= -0.92 \pm 0.08$ coming from combined analysis of low-redshift data.
    Several works use the $H_0$ tension to strengthen the evidence either for a DE component or for modified gravity \citep{Sahni14,Ding15,Zhao17a}.
    On the contrary, \citet{Marra13} show how the inhomogeneities in the cosmic metric can produce a higher value of local expansion rate, while \citet{Wojtak17} claim the higher $H_0$ value is expected in a model with redshift remapping.
    
    In every cosmological model the age of the Universe, $t_0$, is directly related to the value of present expansion rate.
    Therefore, in \Cref{fig:isoch} we also plot the three isochrones that would correspond to the cosmic ages of 13.2, 13.7, and 14.2 Gyr in the standard model.
    There is a quite good convergence to the range of $t_0=(13.7\pm0.5)$ Gyr coming from absolute ages of stellar systems and different classes of observations \citep{Freedman10,Bono10,Monelli15}.
    The age predictions of our and P16 best-fit $\Lambda$CDM models perfectly agree with these, while the R18 value is pulling towards a slightly younger Universe.

  \section{Summary}
    \label{sec:con}
We have utilised the most recent low-redshift data to test the cosmological models, specifically to access the current state of discrepancy in the estimate of the present expansion rate, evidence for acceleration, and phenomenological study of DE component. We find that the high-redshift CMB estimate of $H_0$ is very well consistent with the low-redshift estimate in the $\Lambda$CDM scenario, while it still remains discrepant with the higher local distance ladder estimate. This combined discrepancy among the three estimates of $H_0$ could indicate a need for better modelling of the universe.

Testing for the present cosmic acceleration using the low-redshift data, we find an evidence of $5.8 \sigma$ from BAO data alone and $8.4 \sigma$ from our joint analysis in the $k\Lambda$CDM scenario. Now that the evidence for acceleration is much more significant and pronounced, it demands for a robust cosmological modelling, and to make progressive steps towards a better understanding of DE. We have subsequently shown that the power-law and the linear coasting cosmologies have been unable to compare with the standard model in utilising the information criteria for same. However, given the arguments for these models and several modifications to the analyses methods that would provide contrary results, they yet demand a need to be tested more robustly to arrive at conclusive results. Given the strong evidence for acceleration the quest remains to test its nature and access the dynamics. Among the several directions to proceed, model- independent analysis and improved phenomenological approaches can provide better insights in to the dynamics that can ease the theoretical modelling.

In our work we find mild deviations from the $\Lambda$CDM in the two parameter extended $kw$CDM and $w_0w_a$CDM scenarios using the most recent BAO data. However, it must be kept in mind that none of the low-redshift data provide stringent need for a curvature-free cosmological model. This could in fact be an indication for a dynamic nature of DE. We also studied a LTB model with a Gaussian profile, which is strongly disfavoured with respect to the concordance model by information criteria, such as AIC/BIC or Bayes factor.

    The upcoming opportunities to obtain more observational data (see e.g. \citep{Laureijs11} and \citep{DESI16}) will pave the way for a more exciting era to study cosmological models.
    Those will increment our ability to test the models to higher degree of precision and allow us to make more informed inferences towards understanding the physics of the components of our universe.

  \begin{acknowledgements}
    We thank Giuseppe Bono for his constructive comments and helpful suggestions.
  \end{acknowledgements}

\bibliographystyle{aa}
\bibliography{references}

\end{document}